\documentclass[useAMS,usenatbib]{mn2e}
\usepackage{times}
\usepackage{epsfig}
\usepackage{subfigure}
\usepackage[T1]{fontenc}
\usepackage{aecompl}
\def\atlas3d{ATLAS$^{\rm 3D}$}
\def\Reff{$R_{\rm e}$}
\def\etal{et.\ al.}
\def\lamR{$\lambda_R$}
\def\PAkin{PA_{\rm kin}}
\def\PAphot{PA_{\rm phot}}

\def\lsim{\mathrel{\hbox{\rlap{\hbox{\lower4pt\hbox{$\sim$}}}\hbox{$<$}}}}
\def\gsim{\mathrel{\hbox{\rlap{\hbox{\lower4pt\hbox{$\sim$}}}\hbox{$>$}}}}

\title[Simulating multiple merger pathways to the central kinematics of early-type galaxies]{Simulating multiple merger pathways to the central kinematics \\ of early-type galaxies}

\author[Moody et al.]{
    Christopher E. ~Moody$^1$\thanks{E-mail:cemoody@ucsc.edu}, Aaron J. ~Romanowsky$^{2,3}$, Thomas J. ~Cox$^4$, 
     \newauthor G. S.~Novak$^5$, Joel R. ~Primack$^1$ \\
	$^1$Department of Physics, University of California, Santa Cruz, 1156 High St., Santa Cruz, CA 95064, USA \\
	$^2$Department of Physics and Astronomy, San Jos\'e State University, One Washington Square, San Jose, CA 95192, USA \\
	$^3$ University of California Observatories, 1156 High Street, Santa Cruz, CA 95064, USA \\
	$^4$ Observatories of the Carnegie Institute of Washington, 813 Santa Barbara St., Pasadena, CA, 91101\\
	$^5$ Observatoire de Paris, LERMA, CNRS, 61 Av de l'Observatoire, 75014, Paris, France \\
}

\begin{document}	
\maketitle

\begin{abstract}
	Two-dimensional integral field surveys such as \atlas3d\ are producing rich observational data sets yielding insights into galaxy formation. These new kinematic observations have highlighted the need to understand the evolutionary mechanisms leading to a spectrum of fast-rotators and slow-rotators in early-type galaxies. We address the formation of slow and fast rotators through a series of controlled, comprehensive hydrodynamical simulations sampling idealized galaxy merger scenarios constructed from model spiral galaxies. Idealized and controlled simulations of this sort complement the more `realistic' cosmological simulations by isolating and analyzing the effects of specific parameters, as we do in this paper.  We recreate minor and major binary mergers, binary merger trees with multiple progenitors, and multiple sequential mergers. Within each of these categories of formation history, we correlate progenitor gas fraction, mass ratio, orbital pericenter, orbital ellipticity, and spin with remnant kinematic properties. We create kinematic profiles of these 95 simulations comparable to \atlas3d\ data. By constructing remnant profiles of the projected specific angular momentum ($\lambda_R= <R|V|> / <R \sqrt{V^2+\sigma^2}>$), triaxiality, and measuring the incidences of kinematic twists and kinematically decoupled cores, we distinguish between varying formation scenarios. 

	We find that binary mergers nearly always form fast rotators. Slow rotators can be formed from zero initial angular momentum configurations and gas-poor mergers, but are not as round as the \atlas3d\ galaxies. Remnants of binary merger trees are triaxial slow rotators. Sequential mergers form round slow rotators that most resemble the \atlas3d\ rotators.

\end{abstract}



\section{Introduction}

Measurements of the kinematics of early-type galaxies (ETGs, ellipticals and lenticulars) have revealed novel properties otherwise undetected by photometry. The earliest kinematic measurements of ETGs hinted toward a new classification: some galaxies rotated slowly and supported their flattened shapes by anisotropy in the velocity dispersion tensor, whereas others were supported by more rapid rotation \citep{1983ApJ...266...41D,1977ApJ...218L..43I,Binney:1985p78,Binney:2005p127}.  These dynamical differences were found to correlate with photometric
properties such as isophote shape (boxy or disky), central luminosity profile (cored or coreless)
and total luminosity \citep{1988A&A...193L...7B,1988A&AS...74..385B,1996ApJ...464L.119K,1997AJ....114.1771F}.  It was therefore suggested that ETGs consisted of two fundamentally different classes of object, with distinct formation mechanisms.

This emerging picture of ETGs was codified with the advent of the SAURON integral-field spectrograph \citep{2001MNRAS.326...23B} and its eventual application to a large, volume-limited sample of nearby galaxies, \atlas3d\ \citep{2011MNRAS.413..813C}.
This work has focused on robust quantification of the stellar specific angular momentum within
an effective radius (\Reff, enclosing half of the projected light), and has shown that ETGs
separate cleanly into two populations: the ``fast'' and ``slow'' rotators \citep{2007MNRAS.379..401E,2011MNRAS.414..888E}.
Fast rotators are by far the most common overall (86\% of the sample) and tend to have regular velocity fields indicative of disc-like, oblate-axisymmetric structure \citep{2011MNRAS.414.2923K}. 
Slow rotators are dominant among the most luminous ETGs, while hosting kinematically distinct regions, counter rotating cores, and generally more complex velocity fields that imply a somewhat triaxial structure with all three axes unequal.

The fast--slow rotator dichotomy has been proposed as a replacement for traditional morphology-based classifications of ETGs \citep{2011MNRAS.414..888E},  raising concerns that focusing on the central data provides a biased view
\citep{Foster:2013,Arnold:2014}. Nonetheless, 
it is now a key goal to decipher the formational pathways for the different kinematical sub-types, and to connect them with the overall quest to understand the assembly of the red sequence of galaxies (e.g.\ \citealt{2006MNRAS.366..499D,2007ApJ...665..265F,2011MNRAS.417..845K}).

The vast majority of theoretical work on ETG kinematics has focused on the classic framework of binary {\it major mergers}: the collision of two comparable-mass spiral galaxies.
Violent relaxation in the merger leads to spheroidal structure in the remnant, along with destruction of the initial kinematic patterns.  However, significant rotation can be retained -- from residual disc structure, from orbital angular momentum transfer, and from re-formation of a disc out of any lingering cold gas.  It is thus relatively easy to simulate the formation of major-merger remnants resembling fast-rotators \citep{1999ApJ...523L.133N,Bendo:2000p35,2006ApJ...650..791C}, with the merger parameters potentially being diagnosed through more detailed kinematic information \citep{2001ApJ...554..291C,2001ApJ...555L..91N,2006MNRAS.372L..78G,2007A&A...476.1179B,Hoffman:2009p68}.

Slow rotators, on the other hand, have been much more challenging to reproduce in simulations.  With gas-poor (``dry'') conditions and near head-on collisions, it is possible to end up with little net rotation, but such remnants are also more elongated, discy and internally anisotropic than the observed slow rotators \citep{2003ApJ...597..893N,2006ApJ...650..791C,Jesseit:2009p103,2008ApJ...685..897B,2011MNRAS.416.1654B,2013arXiv1311.0284N}. In cosmological contexts, simulations of dry mergers in groups predominantly yield slowly rotating ($v/\sigma<0.1$) and flattened ($\epsilon > 0.2$) remnants \citep{2013ApJ...778...61T}. 

An encouraging alternative to the major-merger dead-end has recently emerged from hydrodynamical simulations of galaxy formation in a full cosmological context \citep{2007ApJ...658..710N}. These simulations formed massive spheroidal galaxies through multiple mergers, particularly with very uneven mass ratios ({\it minor mergers}).  This scenario is now widely considered as the most likely route to forming the most massive ellipticals, owing in particular to size evolution considerations \citep{2009ApJ...699L.178N,2009ApJ...697.1290B,2012ApJ...744...63O}.
Remarkably, the Naab~\etal\ simulations naturally reproduce the detailed central properties
of slow rotators  -- not only in their kinematics but also in their round, isotropic structure (\citealt{2008ApJ...685..897B}, with similar hints in \citealt{2009A&A...497...35G}).  This outcome is presumably linked to random infall directions for the multiple mergers \citep{Vitvitska:2002p89,2006MNRAS.370.1905H,2012ApJS..203...17R}, but the driving mechanisms have not actually been isolated, for example with an understanding of the relative impacts of major and minor mergers.
Also, although there is a rich history of dedicated simulations of multiple and minor mergers
(e.g. \citealt{Barnes85,1996ApJ...460..101W,Bekki01,2007A&A...476.1179B}), none of them has directly 
engaged with the recent observational constraints on slow, round rotators.

Our paper fills in these missing links by carrying out a suite of controlled merger simulations that is unprecedented in scope, including both major and minor mergers, in both binary and multiple configurations.  By comparing the merger remnant properties to observational trends from \atlas3d, we hope to systematically identify plausible pathways to forming both fast- and slow-rotating ETGs.

The outline of this paper is as follows.
Section~\ref{sec:sim} describes the merger simulations, including their initial conditions and parameter space setup. 
Section~\ref{sec:rem} presents the analysis methodology, keeping in mind direct comparisons to observational efforts. 
Section~\ref{sec:cor} presents correlations with the initial conditions, including orbital parameters and progenitor properties, to their locations on the \lamR--$\epsilon$ and \lamR--$T$ diagrams. 
Section~\ref{sec:kin} makes a similar set of analyses, but instead of correlating to \lamR--$\epsilon$ we attempt to find specific initial conditions leading to kinematically decoupled cores (KDCs) and kinematic twists (KTs).
In Section~\ref{sec:sum} we summarize our findings.


\section{Merger simulations}\label{sec:sim}
Our analysis will focus on 95 simulated remnants of galactic mergers, covering a wide variety of initial conditions as will be discussed below.  Many aspects of the simulations and the remnants were detailed in
\cite{2004ApJ...607L..87C,2006ApJ...650..791C,2008MNRAS.384..386C}, with the novelties
here including an extension to multiple merger histories (initially discussed in \citealt{2008PhDT.........6N}) and an expanded kinematics analysis.

In brief, the simulations use the numerical $N$-body / smoothed-particle hydrodynamics (SPH) code \textsc{gadget} \citep{Springel:2001p59} with entropy conservation enabled. Star and dark matter particles are collisionless and subject only to gravitational forces, while gas particles also experience hydrodynamical forces. The gravitational softening lengths for stars and dark matter are 100 and 400~pc, respectively, while for the gas, the SPH smoothing length is required to be greater than 50~pc. Gas particles are transformed into star particles following the observed rates between gas and star formation rate surface densities. The gas has a `stiff' equation of state where the pressure is proportional to the square of the density.
The star formation begins above a density threshold of 0.0171\,$M_\odot$\,pc$^{-3}$ with an efficiency of 3 per~cent.
Feedback from supernovae is also included, and acts to pressurize the interstellar medium and regulate star formation;
it is implemented as a gradual transfer of turbulent energy to thermal pressure support on 8~Myr timescales.
The details of these recipes can be found in \citet[Table 1 in particular]{2006MNRAS.373.1013C}.
An active galactic nucleus is not included.

Simulations of binary mergers under a variety initial conditions have not generally found a prescription to create slow and round rotators. As a result, we experiment with two types of multiple merger histories. These simulations are not designed to have realistic merger histories, in contrast to formation histories that semi-analytic models or cosmological simulations might produce. Instead, these cases are idealized, featuring progenitors modeled after local galaxies but with varying initial positions. Progenitor spin orientations are randomly chosen over the unit sphere, and eccentricities are chosen between 1.0 and 0.95. Binary merger trees feature progenitors that build mass exclusively in repeated, equal-mass mergers. While binary merger tree simulations test a formation history of extended major-merging, sequential mergers remnants are formed from decreasing mass ratio mergers. Sequential mergers start with an initial merger equal in mass, but as the remnant builds up mass, subsequent identical mergers become increasingly more minor. Both scenarios are simulated with three initial pericenters of 1.5 kpc, 3.0 kpc, and 6.0 kpc, keeping in mind that higher pericenters correspond to increased initial orbital angular momentum. While the magnitude of the orbital angular momentum is systemically varied, the progenitors' angular momenta orientations are randomized, and thus the relation between the internal angular momentum and the initial angular momentum is obscured. Instead, we focus on the effect of increasing the number of progenitors and varying the merging scheme.

In all of our simulations, we use twelve unique spiral galaxies as the building blocks and construct varying formation histories. Each progenitor galaxy is composed of a disc of stars and gas, a stellar bulge and a dark matter halo. We base the galaxies on isolated low-redshift systems, rather than attempting to evolve them from initial cosmological conditions. 
Table~\ref{table_progenitors} summarizes the progenitor properties.
`G' series galaxies are designed to replicate the properties of observed local galaxies in the Sloan Digital Sky survey (SDSS), and cover a range of masses, gas fractions, and bulge-to-disk ratios. `Sbc' series galaxies are modeled after local Sbc-type spirals with small bulges and high gas fractions, and cover less variation in galaxy parameters than the G series, but a larger variety of orbits. 
The number of collisionless particles per galaxy ranges from 40,000 (G0) to 190,000 (G3), and the dark matter particle masses range
from $1.7\times10^6 M_\odot$ to $9.7\times10^6 M_\odot$, and the star particle masses from $2\times10^4 M_\odot$ to $1.3\times10^6 M_\odot$.
There are 10,000 to 50,000 gas particles per galaxy, with particle masses from $6\times10^4 M_\odot$ to $1.8\times10^6 M_\odot$.

\begin{table}
  \begin{center} 
  \caption{Progenitor galaxy properties, grouped by series.
      $M_{\rm tot}$ represents the total gas, stellar and dark matter mass. $M_{\rm baryon}$ is the sum of gas and stellar masses. $f_{\rm gas}$ is the mass of stars divided by the total baryon mass. $B/D$ is the initial stellar bulge-to-disk ratio. $R_{1/2}$ is half-mass radius of the progenitor galaxy. }\label{table_progenitors}
    \leavevmode
    \begin{tabular}{lrrllll} \hline               
		Type & $M_{\rm tot}$  & $M_{\rm baryon}$ & $f_{\rm gas}$ & $B/D$ & $R_{1/2}$\\ 
		 & $10^{10}M_\odot$ & $10^{10}M_\odot$ &  &  & (kpc)\\ 
		\hline
		\multicolumn{3}{l}{Sbc Series}\\ 
		\ \ \ Sbc & 91 & 10.28 & 0.52 & 0.26 & 7.15\\ 
		\multicolumn{3}{l}{G Series}\\ 
		\ \ \ G0 & 5 & 0.44 & 0.38 & 0.02 & 1.84\\ 
		\ \ \ G1 & 20  & 0.70 & 0.29 & 0.06 & 2.33\\ 
		\ \ \ G2 & 51  & 2.00 & 0.23 & 0.11 & 2.90\\ 
		\ \ \ G3 & 116 & 6.20 & 0.20 & 0.22 & 3.90\\ 
		\multicolumn{3}{l}{G3 gas fraction series}\\ 
		\ \ \ G3gf1 & 116  & 3.09 & 0.43 & 0.32 & 3.49\\ 
		\ \ \ G3gf2 & 116  & 4.18 & 0.58 & 0.52 & 2.89\\ 
		\ \ \ G3gf3 & 116  & 5.40 & 0.76 & 1.34 & 1.77\\ 
		\ \ \ G3gf4 & 116  & 0.68 & 0.10 & 0.20 & 3.96 \\
    \end{tabular}
  \end{center}
\end{table}

Using the library of model progenitors, we construct binary and multiple mergers.  The binary mergers have a range of stellar-mass ratios of 1.6 to 8.9 (total mass ratios of 2.3 to 10.2).  We define major and minor mergers as those with stellar-mass ratios less than and greater than 3.0, respectively.
By varying the orbital eccentricity, pericentre and orbital orientations, we generate multiple initial conditions for each pair of identical galaxies.  These parameters are presented in more detail in Table 2 of \citet{2008PhDT........13C}; they are motivated by orbits in a cosmological context \citep{2006A&A...445..403K} but are not designed to be a statistical ensemble.

We label each merger case according to the progenitor type, followed by a string identifying the unique initial conditions for the simulation. For example, G3G1R is a binary minor merger between the most-massive G3 galaxy and the third-most massive galaxy G1, with the minor galaxy having a retrograde spin. `Sbc' galaxy simulations are always equal mass mergers with two identical Sbc galaxies, and so for brevity we drop the second `Sbc' from the label.

We define major mergers as approximately equal-mass mergers of identical progenitor galaxies, and minor mergers as being between any two distinct galaxies with a mass ratio of more than 3:1. 

\begin{figure}
\includegraphics[width=\columnwidth]{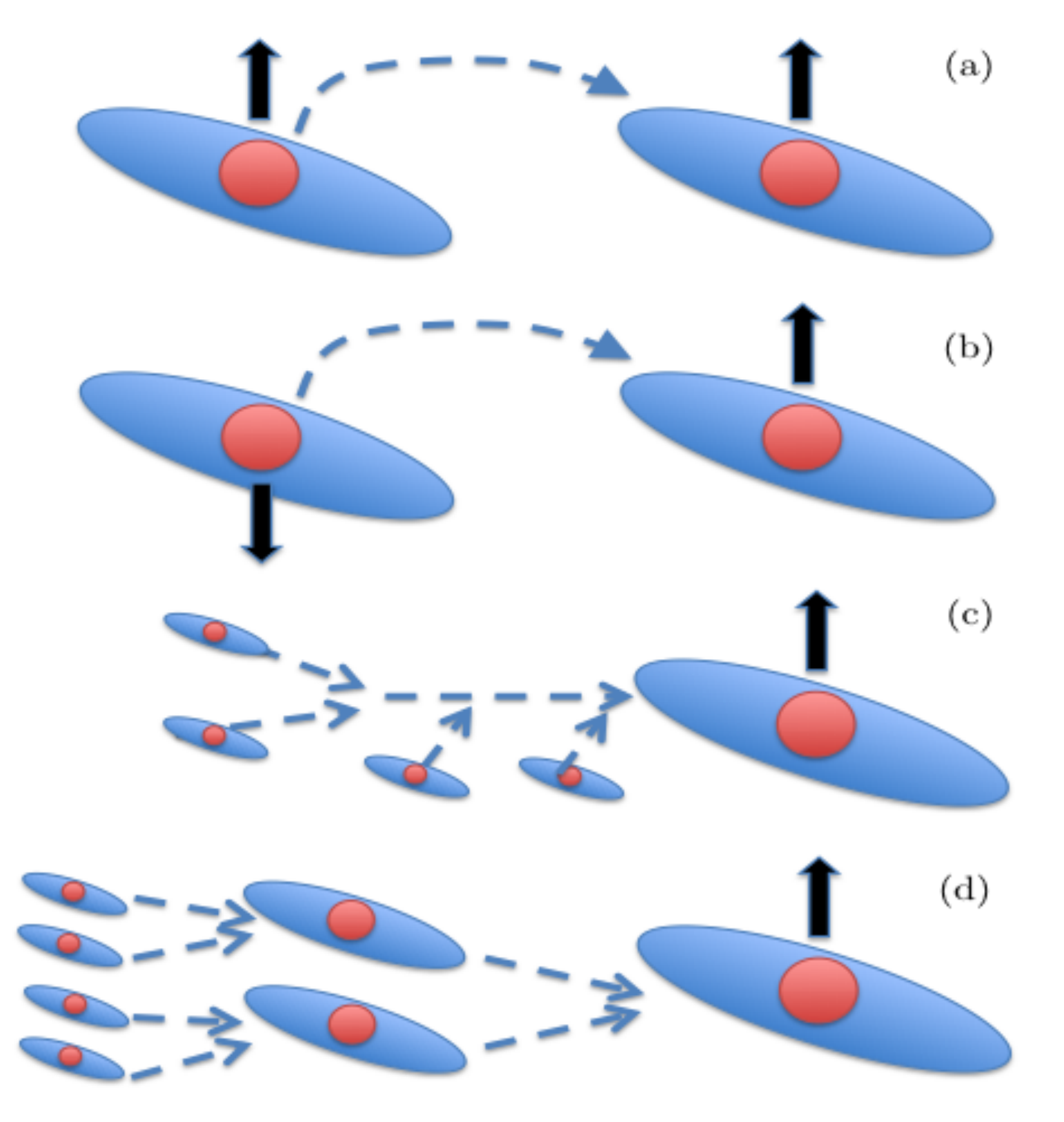}
	\caption{A schematic of the varying orbital orientations and initial conditions of the merger simulations. In all cases, the orbital angular momentum is oriented such that it points upwards. Shown are (a) a prograde--prograde binary merger with the spin of both galaxies aligned, (b) a prograde--retrograde binary merger, (c) a series of sequential mergers and (d) a binary merger tree simulation with each generation of galaxies equal in mass. \label{orbital_orientations}}
\end{figure}

The multiple mergers are constructed in two distinct scenarios that are dominated by major or minor mergers, and are thus designed to bracket qualitatively the more diverse range of histories that would be expected in a full cosmological context.  These scenarios are shown schematically in Figure~\ref{orbital_orientations} and are:

\begin{enumerate}
	\item \textit{Sequential} histories that grow by mergers of four or eight identical progenitors,
	\item \textit{Binary Merger Tree} histories that grow by successive 1:1 mergers.  This models the growth of a galaxy exclusively through multiple major mergers.
\end{enumerate}

The multiple mergers are always constructed from a set of initially identical galaxies, so the labeling (Table~\ref{table_multiple}) simply indicates the number and initial orientations. In such a case, G2-8s-1 indicates that eight G2 galaxies merged sequentially with pericenters of 1.0 kpc. 
In all cases, the progenitors accrete isotropically and at a regular time interval onto the larger galaxy.

\begin{table}
	\caption{ Assembly histories of multiple merger models.
    }\label{table_multiple}
 	\begin{minipage}{140mm}
		\begin{tabular}{|l|l|l|l|l|}
			Name & Pericenter & Number of  & Number of   \\
			 & (kpc) & G2 Progenitors & G1 Progenitors
			\\  \hline
			\multicolumn{5}{|l|}{Sequential series}\\
			G1-8s   & 3.0  & 0 & 8 \\
			G1-8s-1 & 1.5  & 0 & 8 \\
			G1-8s-3 & 6.0  & 0 & 8 \\
			G2-4s   & 3.0  & 4 & 0\\
			G2-4s-1 & 1.5  & 4 & 0\\
			G2-4s-3 & 6.0  & 4 & 0\\
			\multicolumn{5}{|l|}{Binary Merger Tree series}\\		
			G1-8b   & 3.0  & 0 & 8 \\
			G1-8b-1 & 1.5  & 0 & 8 \\
			G1-8b-3 & 6.0  & 0 & 8 \\
			G2-4b   & 3.0  & 4 & 0 \\
			G2-4b-1 & 1.5  & 4 & 0\\
			G2-4b-3 & 6.0  & 4 & 0\\
			
		\end{tabular}
	\end{minipage}

\end{table}


\section{Merger Remnant Analysis}\label{sec:rem}

The goal of this paper is to explore correlations between the merger histories of simulated galaxies and their stellar kinematic properties. These properties are constructed to mimic observational measurements of line-of-sight (LOS) velocity distributions. We focus on a particular metric of specific angular momentum that has been extensively used by the SAURON project and the \atlas3d\ survey:
\begin{equation}\label{eqn:lamR}
	\lambda_R=\frac{\langle R|V|\rangle}{R\sqrt{V^2+\sigma^2}}=
	\frac{\sum_{i}^{N_p}M_iR_i|V_i|}{\sum_{i}^{N_p}M_iR_i\sqrt{V_i^2+\sigma_i^2}} ,
\end{equation}
where $R$ is the projected radius, $V$ and $\sigma$ are the local mean velocity and velocity dispersions, respectively, and $N_p$ spatial bins are designated by index $i$
\citep{2007MNRAS.379..401E}.
For the simulations, the sum is naturally weighted by projected mass $M$ rather than luminosity $L$ as in the observations. The binning is carried out over an elliptical region within the projected half-mass radius \Reff, for direct comparisons with \atlas3d\ observational results.

The lower limit for this angular momentum metric is $\lambda_R=0$, which could correspond to a purely pressure-supported system with no rotation, or to a system where the angular momentum vector is exactly along the line of sight so the rotation is not observable.
The upper limit is $\lambda_R=1$ for pure rotation in a dynamically cold disk.
The rough boundary between slow and fast rotators is $\lambda_R\sim 0.1$, with a more precise diagnostic to be discussed in Section~\ref{sec:cor}.


\begin{figure}
	\psfig{file=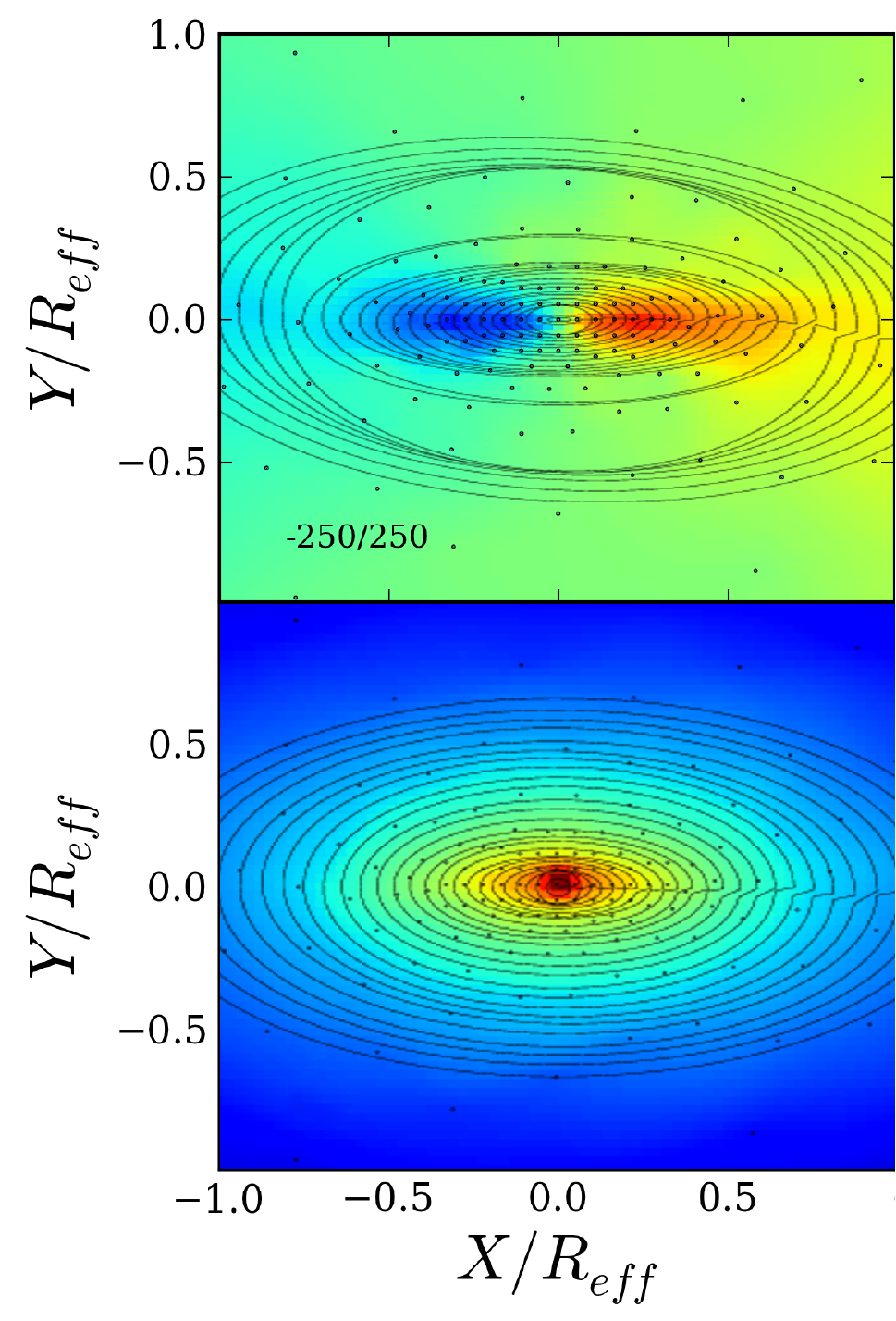,height=6.0cm}
	\psfig{file=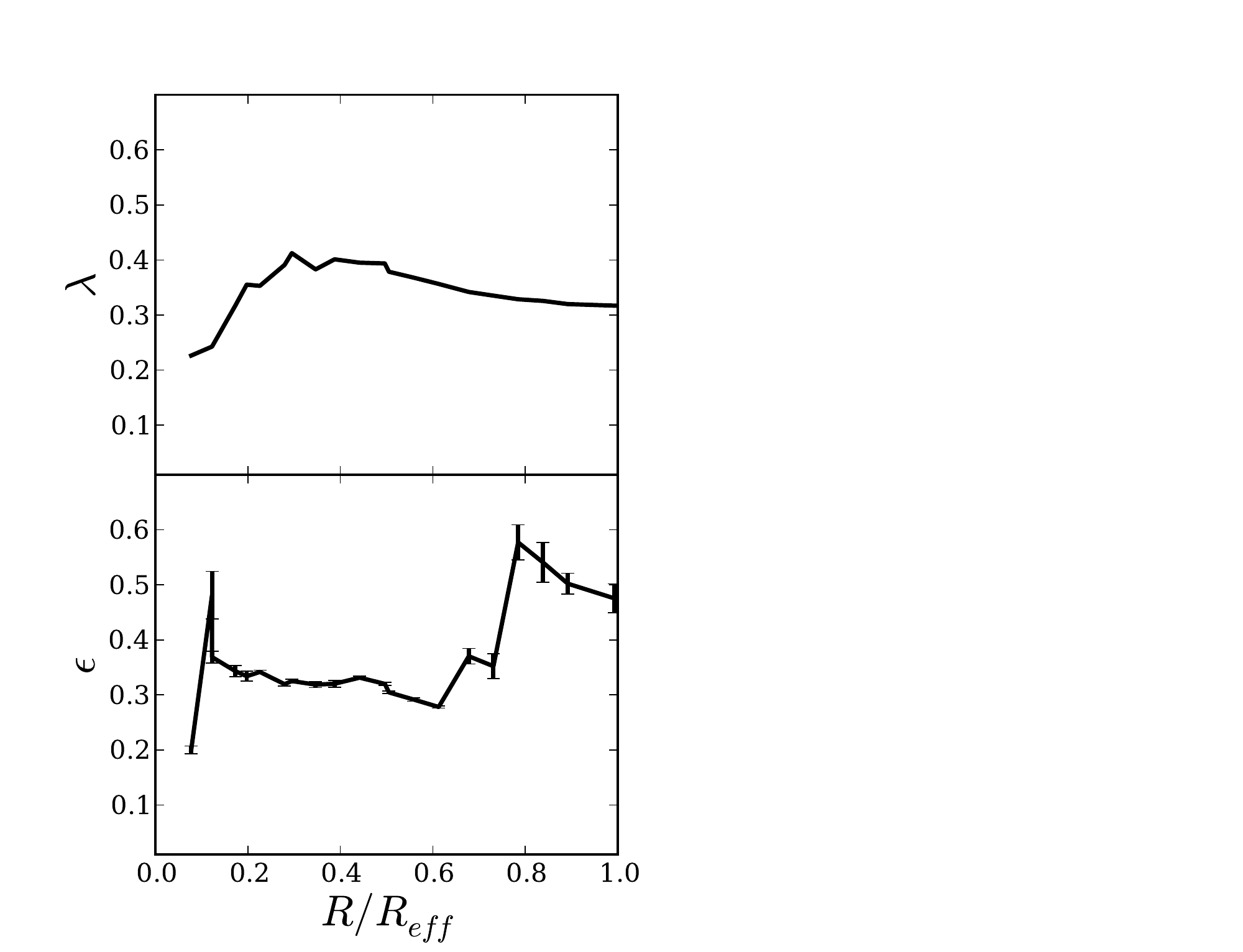,height=6.5cm}
	\caption{The edge-on (top left) line-of-sight velocity and (bottom left) density of the fiducial remnant, SbcPP. The \lamR\ parameter vs. radius is shown in the top right and the ellipticity of the kinemetric ellipses is shown in the bottom right. Black dots are the Voronoi centroids that ensure a minimum signal-to-noise ratio. As an example, the kinemetric and photometric best-fitting ellipses are overlaid. \label{example}}
\end{figure}

As the first step in calculating \lamR, for every merger remnant we project the stellar component, including both the initial stars and those formed during the course of the simulation
(Figure~\ref{example}).
We use 128 viewing angles randomly chosen on a sphere. Once projected, we bin particles in two spatial dimensions with sides of $\approx 0.03$~\Reff. We then use an adaptive Voronoi binning software to ensure that a minimum of $\sqrt{N}\geq20$ particles are enclosed in each bin \citep{2003MNRAS.342..345C}. This binning scheme combines nearby bins with few particles to create an aggregate bin with a higher signal-to-noise ratio, while keeping the region as compact and as uniform as possible.  This adaptive binning is particularly useful beyond $0.5$~\Reff, where the number of particles per bin falls quickly. The centroid of each Voronoi bin is then assigned values of $M_i$, $V_i$, and $\sigma_i$.
We assume these quantities are Poisson distributed, and assign uncertainties proportional to the enclosed number of particles, $\sqrt{N}$.

Given these Voronoi bins, we could proceed to calculate \lamR\ as in equation~\ref{eqn:lamR}.  However, there is a complication that estimates of \lamR\ can be biased high from small-scale fluctuations in numerical simulations, since only the absolute value of $V$ contributes to the calculation. As emphasized by \citet{2010MNRAS.406.2405B}, this problem can lead to slow rotators being misclassified as fast rotators.  We address this problem by calculating \lamR\ from smoothed kinematic fields, whose construction will be discussed below. We find that with standard resolution, noise fluctuations do indeed drive up the values of \lamR\ by $\approx 0.1$ (which would make slow--fast distinctions very problematic) while high resolution cases are biased by only $\approx 0.01$.  Fortunately, the smoothed standard-resolution case turns out to have \lamR\ comparable to the high resolution case, and we conclude that we can reliably identify slow rotators if present among our merger remnants.

We next carry out a Fourier expansion analysis of the remnants called ``kinemetry'' that is needed for several purposes, including generating smoothed kinematic maps, estimating stellar density parameters, and calculating various detailed kinematic properties. The method follows \citet{2006MNRAS.366..787K} and makes use of the public software package \textsc{Kinemetry}. The basic idea is that concentric ellipses are fitted to the Voronoi-binned velocity moments. These ellipses are defined as:
\begin{eqnarray}
	K(\psi)=A_0+A_1 \sin(\psi) + A_2 \cos(\psi) + \nonumber \\
		A_3 \sin(2\psi)+ A_4 \cos(2\psi)+A_5 \sin(3\psi)+A_6 \cos(3\psi) ,
\end{eqnarray} 
with $\psi$ as the angle along an ellipse and the coefficients $A_i$ the amplitude of the photometric or kinematic moments. The best fitting ellipse minimizes $\chi^2=\Sigma_{i=0}^{6}A_i^2$.

We calculate multiple best-fit ellipses for every projected density and velocity map, constraining the long axis of each ellipse to 25 logarithmically-spaced values from 0 to $1.5$~\Reff, and fixing the centroid for each ellipse. From these ellipses we can calculate the position angle ($PA$) and ellipticity, $\epsilon$, as a function of the projected ellipse radius. The ellipticity is defined as $\epsilon=1-b/a$, so that a circle has $\epsilon=0$ and $\epsilon=1$ is an infinitely elongated ellipse. The misalignment angle, $\Psi$ is calculated as the difference between the photometric and kinematic ellipses, $\sin\Psi= |\sin(\PAkin-\PAphot)|$. 

The \lamR--radius profile for a typical edge-on early-type galaxy starts at nearly zero at the center where it is dispersion-supported. As the dispersion support falls off with increasing radius, $\lambda$ climbs roughly in proportion to $V/\sigma$, and ultimately plateaus, largely reflecting the rising rotation curve at larger radii. Barring kinematically decoupled cores (KDCs), twists (KTs), misalignments or other phenomena discussed later, face-on projections tend to show little rotational support and so have \lamR\ values closer to zero.

\begin{figure*}
	\psfig{file=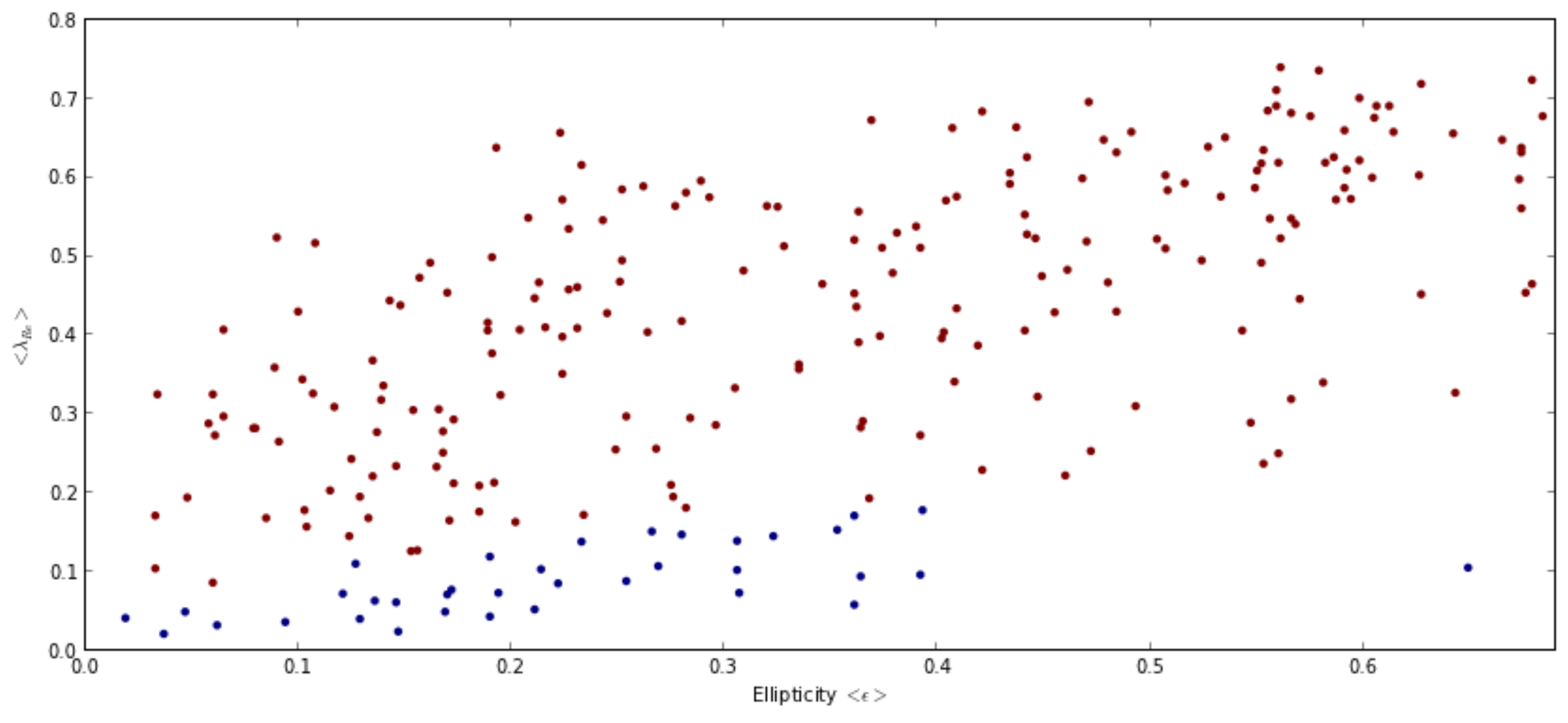,width=15cm}
	\psfig{file=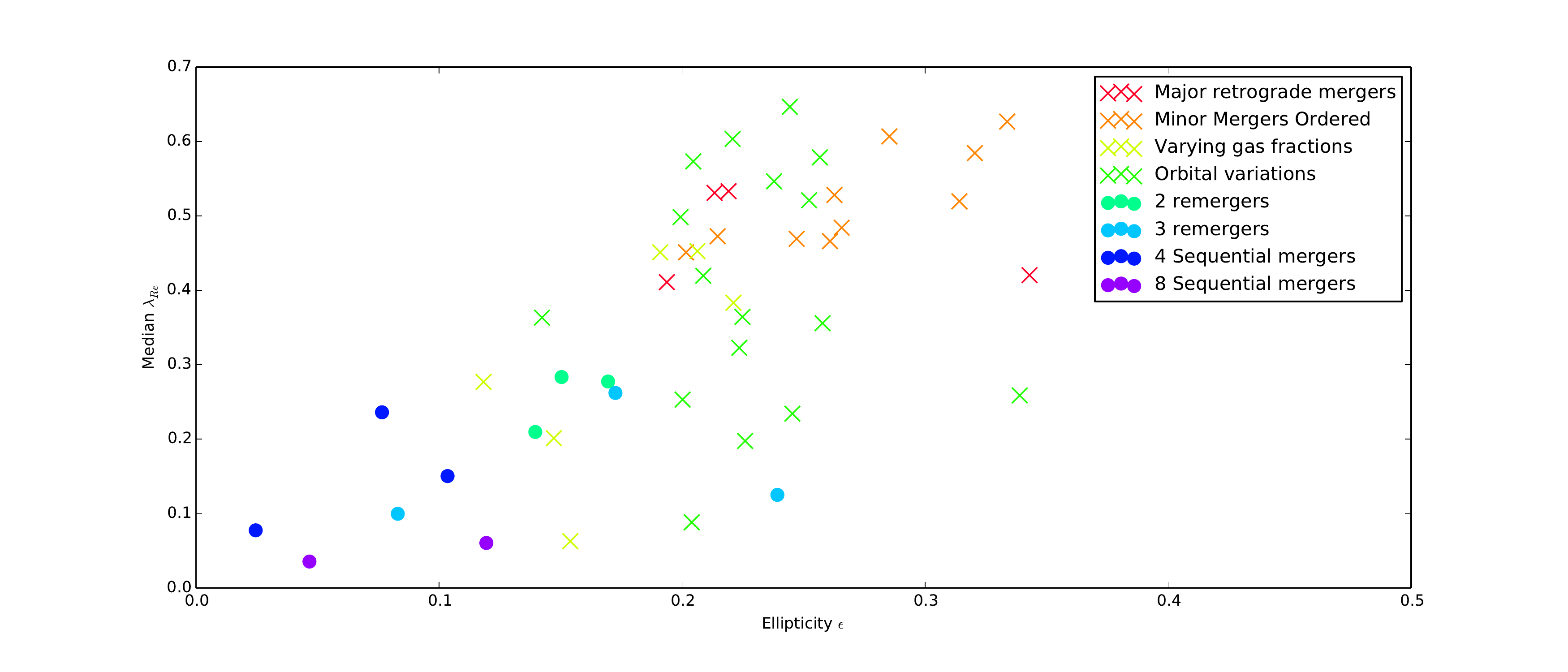,width=18cm}
	\caption{Diagnostic plot of central specific angular momentum versus ellipticity for early-type galaxies, where the dashed curve separates fast and slow rotators.
	The top panel shows the observations from \atlas3d.
	The bottom panel shows simulated merger remnants where each point represents a different case (generally classified according to the legend). 
	The point location is given by the median \lamR\ over many projection angles, and the corresponding ellipticity, $\epsilon$ for that point.
	All classes of binary mergers are generally consistent with the observed family of fast rotators, while only the sequential, multiple mergers match the observed round, slow rotators.
	}
	\label{lamellip}
\end{figure*}

As a measure of unordered rotation we use the normalized amplitude of the first terms not minimized by the best fit procedure, specifically the fifth sine and cosine terms,     

\begin{equation} \frac{k_5}{k_1}=\frac{\sqrt{A_5^2+B_5^2}}{\sqrt{A_1^2+B_1^2}} \end{equation}

$k_5$ is analogous to disciness or boxiness in isophotes, and represents deviations from simple rotation. The mean uncertainty in the observed SAURON data is roughly $\frac{k_{5}}{k_{1}}\sim0.015$. Smaller values are consistent with regular rotation, but deviations from regular rotation tend to increase $k_{5}$.

In section \ref{sec:kin} we discuss the correlation of triaxiality and projected ellipticities with \lamR. To quantify the shapes of merger remnants we iteratively diagonalize the moment of inertia tensor using an ellipsoidal window \citep{2006ApJ...646L...9N}. The eigenvectors of the tensor yield the ellipsoid axes which in turn can be used to calculate the three-dimensional triaxiality $T=1.0-(a^2-b^2)/(a^2-c^2)$. Ellipsoids are oblate, triaxial, or prolate if $T > 0.75$, $0.25 \leq  T \leq 0.75$, or $T < 0.25$ respectively.

The majority of the kinematic maps are dominated by regular rotation, where $\overline{k_5/k_1}$ tends to be smaller than 0.04. For the rest of the maps, we adopt the kinematic classification scheme of \citet{2011MNRAS.414.2923K} and find the rate of kinematic twists (KTs) and kinematically decoupled cores (KDCs) in each LOS velocity map. Kinematic twists are defined are defined to smoothly change by at least a $10^{\circ}$ in $\PAkin$ over the whole map. Kinematically decoupled cores are defined by abrupt changes in $\PAkin$ of at least $30^{\circ}$ in a region of zero-velocity.


\section{Parameter study of remnant properties}\label{sec:cor}

We now focus on two key diagnostics for characterizing galaxies.
The first is the classification as fast or slow rotators, as developed from empirical analysis of \atlas3d\ data by \citet{2011MNRAS.414..888E}.
It is based on two parameters:  the specific angular momentum \lamR,
and the projected ellipticity $\epsilon$.  The boundary between slow and fast rotators in the \lamR--$\epsilon$ plane for
a 1~\Reff\ aperture is:
\begin{equation}
\lambda_R=0.31\sqrt{\epsilon} .
\end{equation}

The second diagnostic is based on the galaxies' observed non-uniform occupancy of the \lamR--$\epsilon$ plane (see first panel of Figure~\ref{lamellip}). The fast rotators occupy a characteristic quasi-diagonal region that is understood to trace a relatively homologous family of near-oblate rotators at random viewing angles. 
We will not discuss this distribution in further detail since fast rotators are not the primary focus of this paper.
The slow rotators are generally very round, with almost all of them having $\epsilon \lsim 0.35$.
This is a fundamental and challenging constraint for formation models of these galaxies, which may be better called ``round slow rotators''.

Here it is important to recognize that it is not enough for the simulations to produce an occasional object with the right apparent properties, since for example, this can happen for a fast rotator viewed nearly face-on.  The goal is for the {\it ensemble} of projections for a given set of merger simulations to resemble the ensemble of observations of slow rotators.

We now carry out a series of parameter studies, where we investigate systemically the effects of each parameter (among the progenitor properties and orbital variations) on the merger remnant in the \lamR--$\epsilon$ plane (other remnant properties will be discussed in the next
Section).  A preview summary of the results is shown in the second panel of Figure~\ref{lamellip}, where the median \lamR\ and the corresponding $\epsilon$ of each type of remnant has been evaluated over all projection angles and coloured by category of initial conditions. Many of the remnants reproduce broadly the properties of the fast rotators, while only a very limited subset match the observed round slow rotators.

We focus on quantifying and separating fast from slow rotators by making extensive use of the \lamR--$\epsilon$ diagnostic diagram. Tests of merger remnants show that \lamR\ is a robust indicator of the angular momentum content and that the confusion rate between a face-on fast rotator and a true slow rotator is small \citep{Jesseit:2009p103}.

\begin{figure}
	\psfig{file=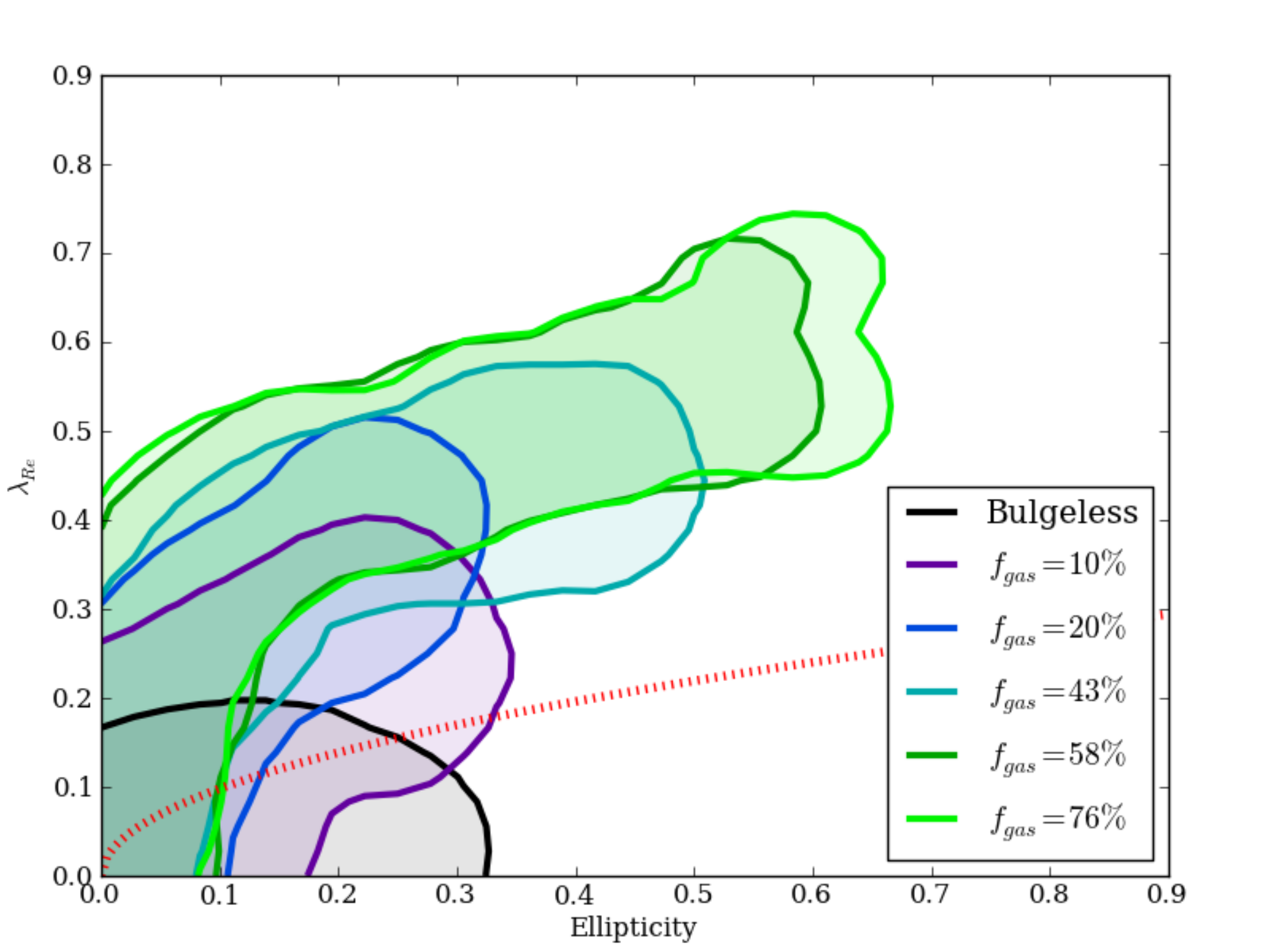,height=7cm}
	
	\caption{Central specific angular momentum versus ellipticity for simulated major merger remnants, for a series of varying gas fraction.  Here each curve outlines all projections for a given simulation, color-coded as in the legend.  Increasing the gas fraction yields faster rotators. In the simulation lacking a bulge, star formation in the progenitor is much higher than in the fiducial case prior to the merger event. The bulgeless case has an effective gas fraction of less than 10\%. Averaged over all viewing angles, the bulgeless case is rotating marginally faster than the \atlas3d\ relation.
	}
	\label{le-gasfractions}
\end{figure}

We begin with binary mergers, where
our first parameter study is of the initial gas fraction $f_{\rm gas}$, using the G3 series of progenitors.  Our fiducial case has $f_{\rm gas}=$~20\%, and we cover a range of 10\% up to 76\%. 
We also try a bulgeless version of G3, which is effectively a dry merger ($f_{\rm gas}\sim$~0\%), with most of the available gas converted into stars before the merger even begins: although this case is not realistic, it provides a bracketing limit for isolating the effects of gas.   
The results are shown in Figure~\ref{le-gasfractions}, where each curved line encloses all projected values for a given simulation (color-coded as in the legend).
We see that each remnant populates a characteristic diagonal region upward to the right in \lamR--$\epsilon$  space. This is because a flattened, rotating system seen edge-on shows the maximum rotation and elongation, while more face-on projections reduce both observed quantities in a correlated fashion.

Marginalizing out the projection effects, we see that
there is a systematic trend for remnants of increasing gas fractions to be faster and more elongated rotators. 
We conclude that the extra gas has the effect of funnelling angular momentum to inside the half-mass radius, increasing rotational support in the interior and shrinking the semimajor axis of the remnant -- which mirrors the results of \citet{2006MNRAS.373.1013C}. 
We note however that this effect could be weakened or even reversed if the gas were treated as clumpy and turbulent, and with earlier star formation, so that the merger
became effectively collisionless (cf.\ \citealt{Teyssier:2010,Bournaud:2011}).
The bulgeless case has $<$\lamR$>=0.15$ and $<\epsilon>=0.10$, placing it marginally above the slow-rotator criterion. Therefore the remnant is rotating just quickly enough to not be a slow rotator.

\begin{figure}
	\psfig{file=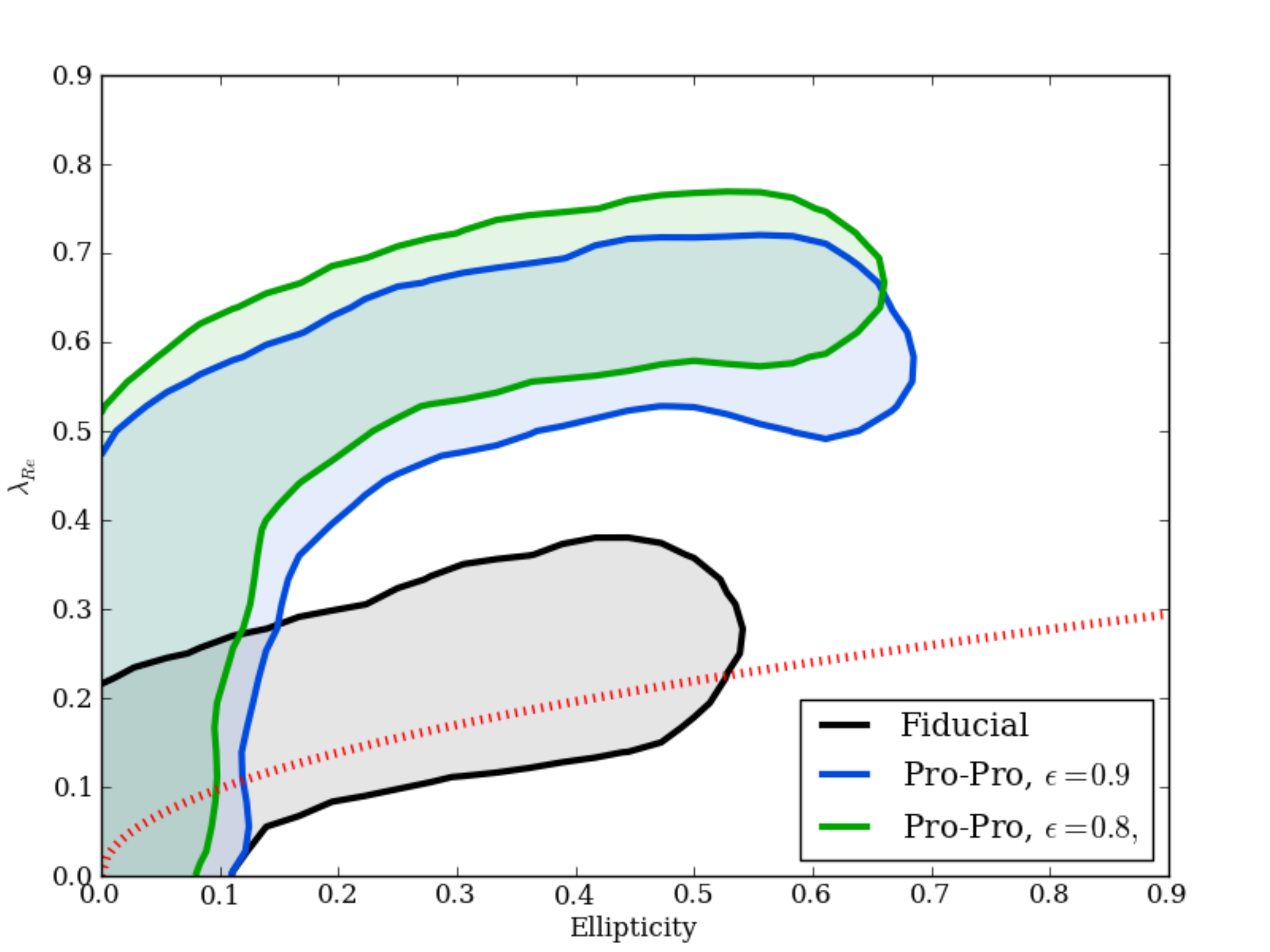,height=7cm}
	\caption{Same as Figure \ref{le-gasfractions}, but with variations of the merger orbits
	to lower eccentricities. These lead to faster rotating remnants.
	}
	\label{le-ellipticity}
\end{figure}

Next we consider lower eccentricities of the merger orbits ($e=0.8$ and $0.9$ versus
$1.0$ for the fiducial case), at fixed pericentric distance, using SbcPP progenitors.
The more circular orbits increase the initially available angular momentum and result in a more rapid merger,
with less time for initial gas consumption and for 
the dark matter halo to absorb  angular momentum.
All of these effects (detailed in \citealt{2006ApJ...650..791C}) 
lead to higher rotational support of the remnant, as shown in Figure~\ref{le-ellipticity}.
 
\begin{figure}
	\psfig{file=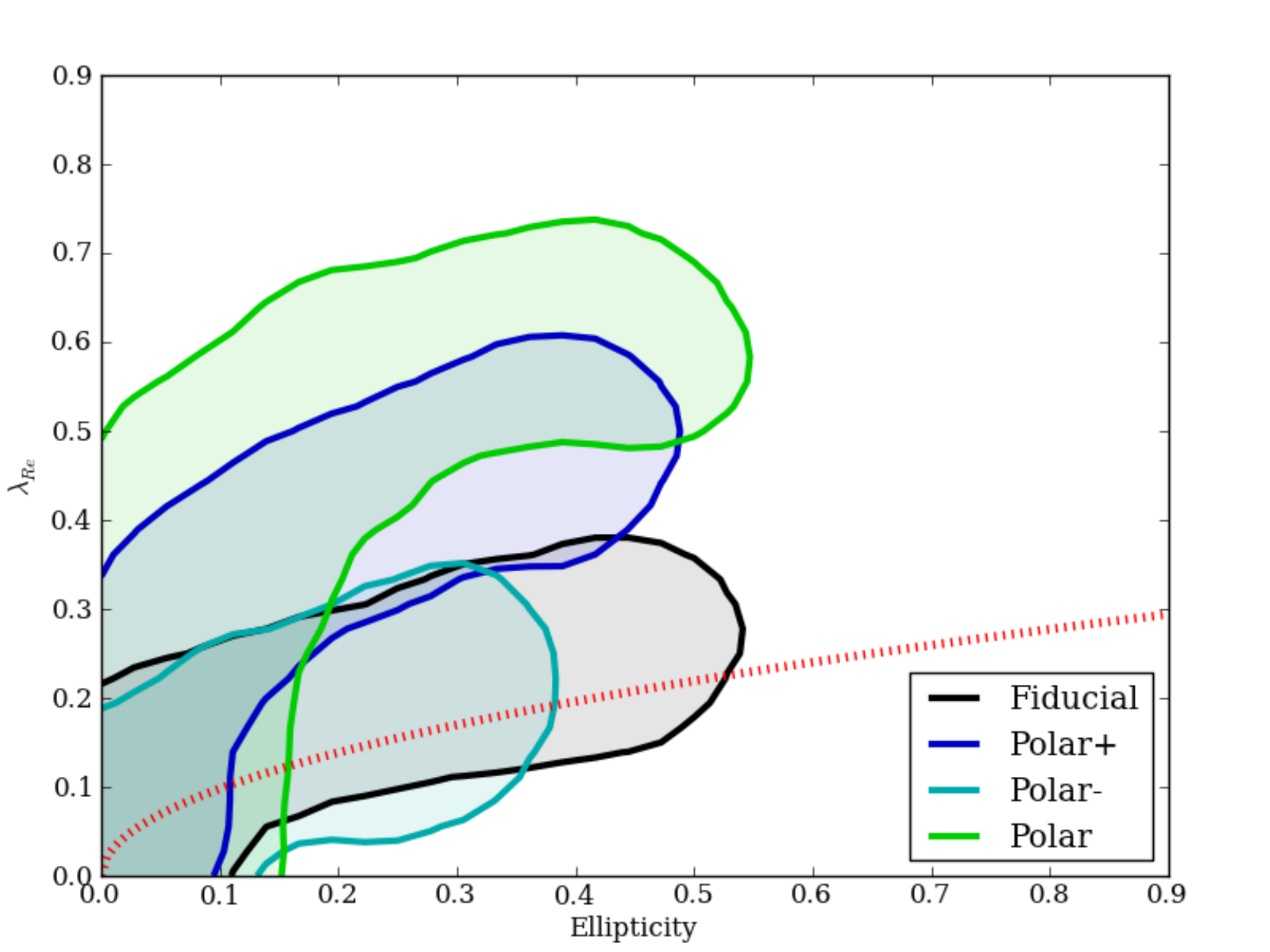,height=7cm}
	\caption{Same as Figure \ref{le-gasfractions}, but showing varying polar orbits in Sbc-type galaxies. Polar orbits vary the orientation of the galactic angular momentum with respect to the orbital angular momentum axis. The `Polar-' case minimizes by construction the angular momentum in any one axis. While some polar orbits do yield remnants that can be significantly as slowly rotating as the fiducial, these remnants are still far more elliptical than \atlas3d\ slow rotators.}
	\label{le-polar}
\end{figure}

Next we consider variations in initial disk spin directions, using Sbc progenitors.
The fiducial case has all three angular momenta (orbital momenta, and the two progenitors' spins) aligned. 
`Polar' has both progenitors' axes perpendicular to the orbital angular momentum.  The `Polar$-$' (`Polar$+$') case has a single progenitor rotated $90^{\circ}$ such that angular momentum spin axis points away (towards) from the other progenitor. 
The results are shown in Figure~\ref{le-polar}.  The `Polar$-$' case is of interest because its initial
orbital and spin angular momenta are approximately equal, which yields a rounder and slower rotating remnant than the fiducial case.
However, it is still too flattened and fast-rotating ( $\epsilon\approx0.3$, 
$\langle\lambda_R\rangle\approx0.2$) to match the observed slow rotators.

\begin{figure}
	\psfig{file=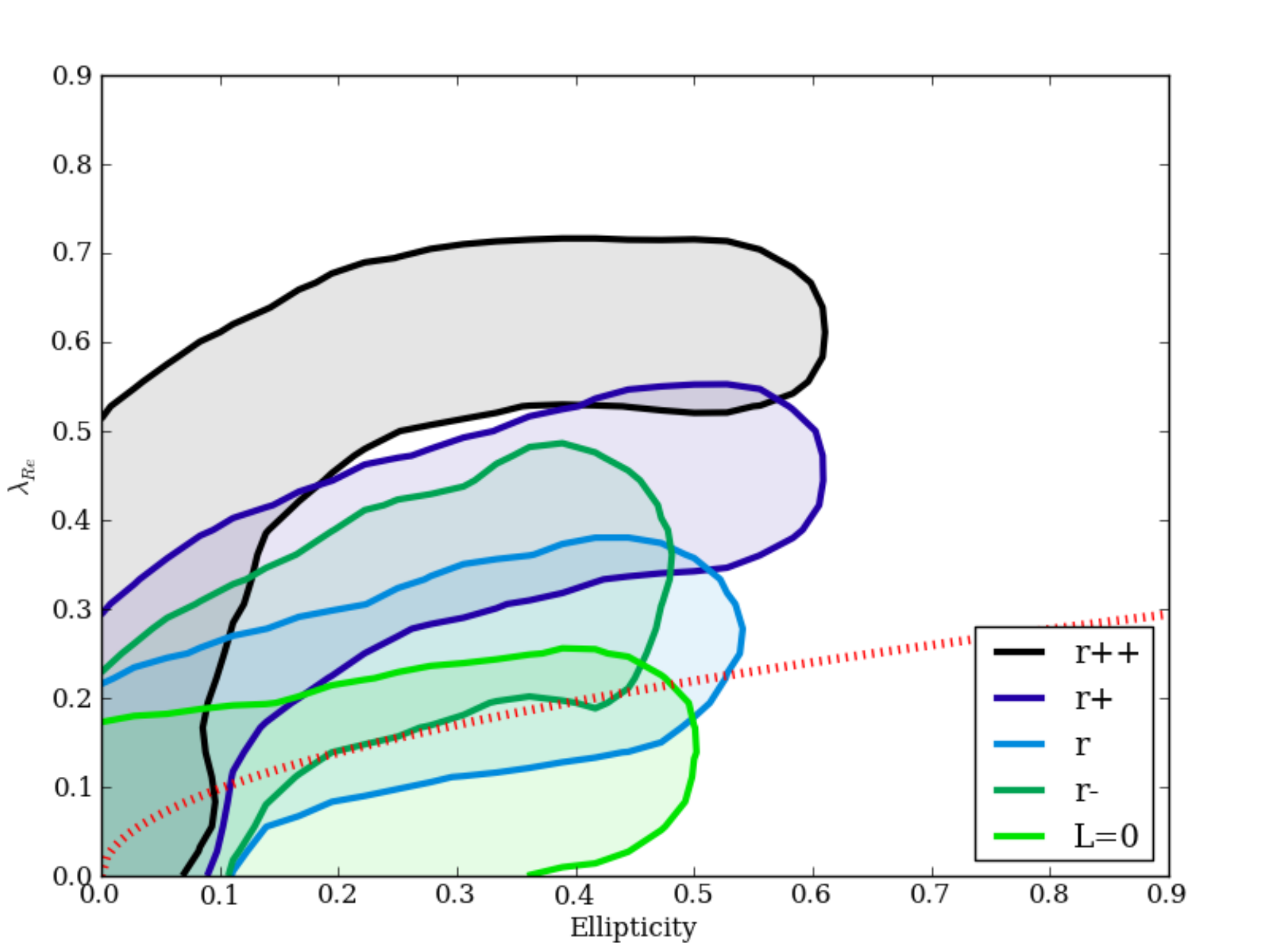,height=7cm}
	\caption{Same as Figure \ref{le-gasfractions}, but showing initial conditions with varying total angular momenta (orbit $+$ spin). As the initial orbital angular momentum is decreased the rotational support of the remnant is imilarly reduced. However, in all cases, the remnant is not as round as \atlas3d\ slow rotators.  
	}
	\label{le-am}
\end{figure}

Variations in the initial orbital angular momentum case are shown in Figure \ref{le-am}. The fiducial case labeled `r' has the standard pericenter distance of 11kpc for SbcPP, a prograde-prograde merger. This fiducial case is compared to a merger where the orbital angular momentum is fixed to negate the progenitor galaxies' spin (labeled `L=0'), thereby constructing a system with nearly zero initial angular momentum. Compared to the fiducial case, the average \lamR\ of the `L=0' remnant is halved, yielding a reduced \lamR\ of about 0.1. Evidently, a zero net initial orbital angular momentum leads to a rotator supported less by rotation than by dispersion. Averaged over all projections, the zero angular momentum case nominally qualifies as a slow rotator. However, the remnant is still highly elliptical ($\epsilon\approx0.4$) and thus not a suitable analogue of \atlas3d\ slow rotators. As the radial pericenter is increased, so the orbital angular momentum rises ($L \propto R^2 $), and so does the amount of rotation in the final remnant. This increased rotation flattens the remnant slightly, with our example simulations increasing from maximum $\epsilon=0.45$ in the zero-angular momentum case to a maximum $\epsilon=0.55$ in the fastest-rotating remnant. The magnitude of the initial angular momentum does not appear to affect the ellipticity of the remnant significantly, a finding shared by \citet{2006ApJ...650..791C} and \citet{2010MNRAS.406.2405B}. So while the rotation of the remnant can certainly be reduced, a baseline flattening due to a dispersion anisotropy remains unchanged.  

This last issue is the overall obstacle to forming the family of slow, round rotators with binary major mergers.
Various effects can modify the rotation (or \lamR),
where we note additionally that the coarse gas treatment in our simulations (stiff equation of state, and minimum resolution of 50~pc) 
could lead to an overprediction of the rotation (see \citealt{2010MNRAS.406.2405B}).
However, none of these merger variations have been shown to produce galaxies that are round enough on average.

\begin{figure}
	\psfig{file=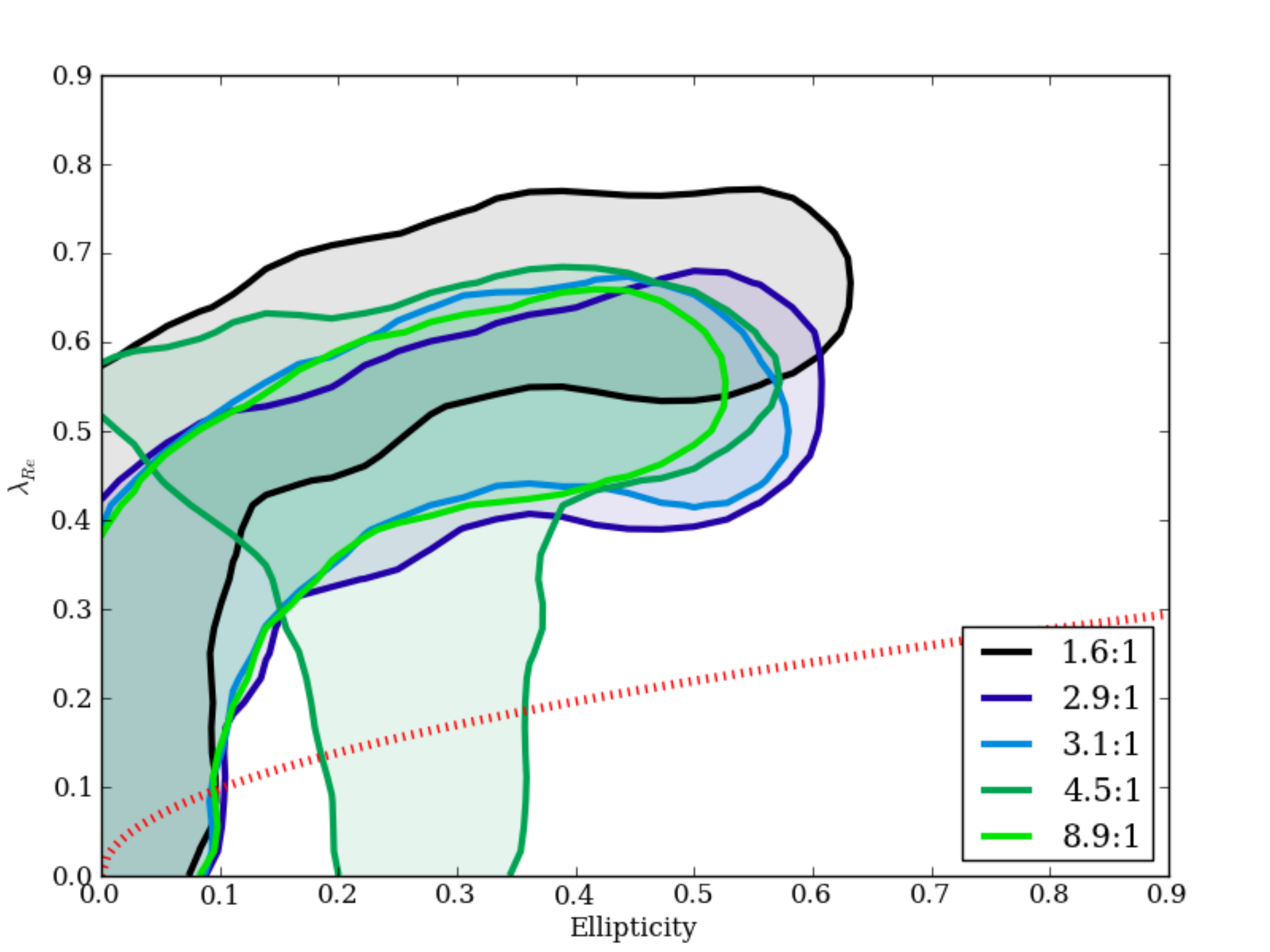,height=7cm}
	\caption{Same as Figure \ref{le-gasfractions}, but showing binary minor merger simulations varying both progenitor masses. Evidently, a single minor merger does not create a slow rotator. The merger ratios presented here range from 1:1.7, G1G0R in black, to 1:10, G3G2R in orange. }
	\label{le-rminor}
\end{figure}

As it is more difficult for minor mergers to disrupt the properties of the larger progenitors, the variation in remnants of individual minor merging is much smaller and less dramatic than in major majors. The mergers shown in Figure \ref{le-rminor} are all retrograde, thereby containing less global angular momentum than the fiducial prograde case. As the merger ratio increases, the minor progenitor negates more of the available initial angular momentum. Since the rotation of the remnant depends on the merger ratio, with more massive progenitors (with smaller mass ratios) leading to higher \lamR\ and slightly more flattening. The effect is not dramatic, with $\lambda_{R, max}$ varying from 0.5-0.7, and $\epsilon$ ranging from 0.4-0.6. 

\begin{figure}
	\psfig{file=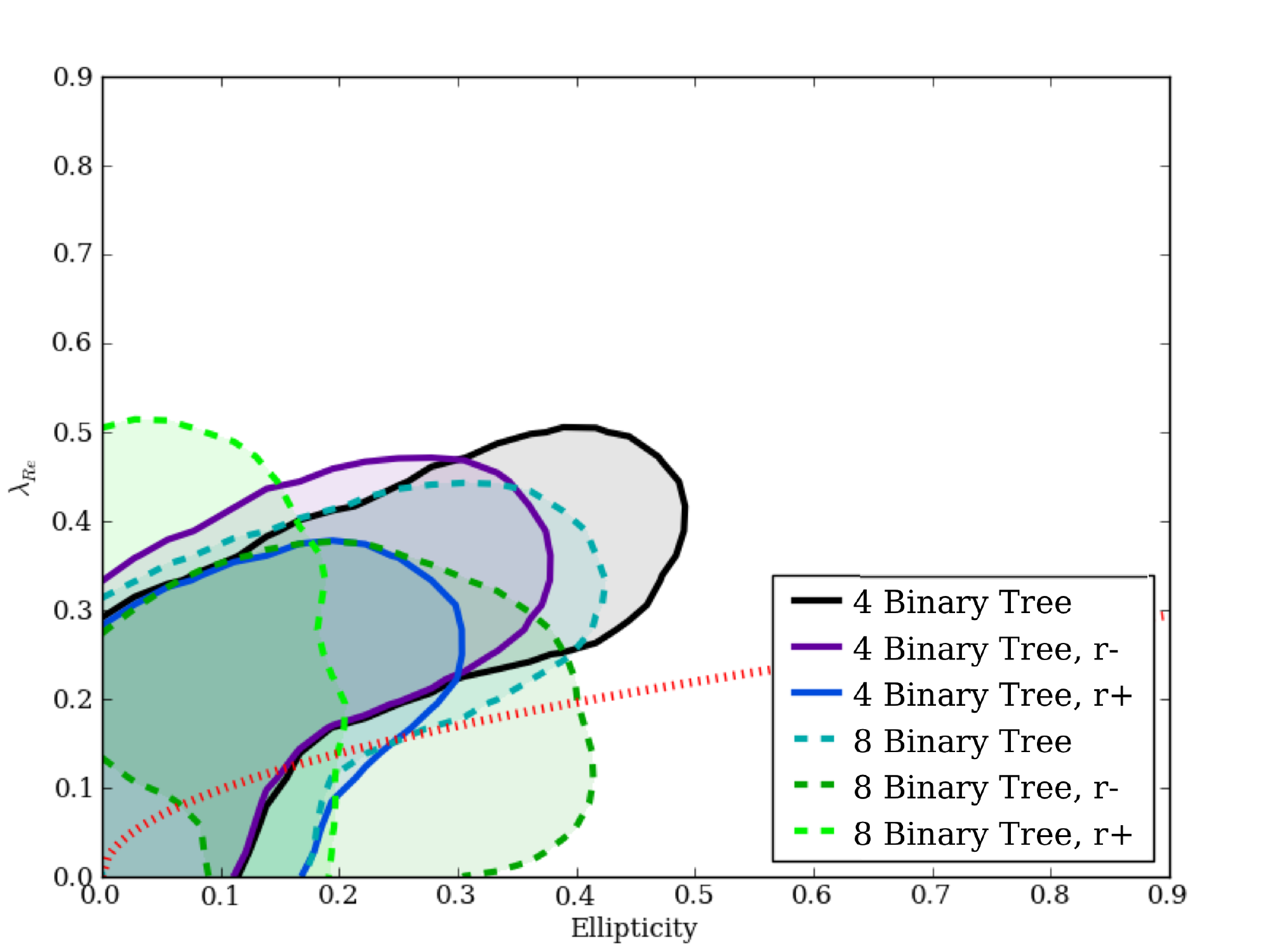,height=7cm}
	\caption{Same as Figure \ref{le-gasfractions}, but showing binary merger tree simulations with either 4 or 8 identical G1 or G2 progenitors. Simulations with four progenitors are shown in solid lines, otherwise eight-progenitor simulations are shown in dashed lines. Binary merger tree simulations of four (eight) progenitors occur in two (three) generations of binary equal-mass mergers. Increasing the number of progenitors in sequential simulations from four to eight yields still rounder and slower remnants, albeit not as slow and round as sequential merger remnants (see Figure \ref{le-sequential}). }
	\label{le-remerger}
\end{figure}

\begin{figure}
	\psfig{file=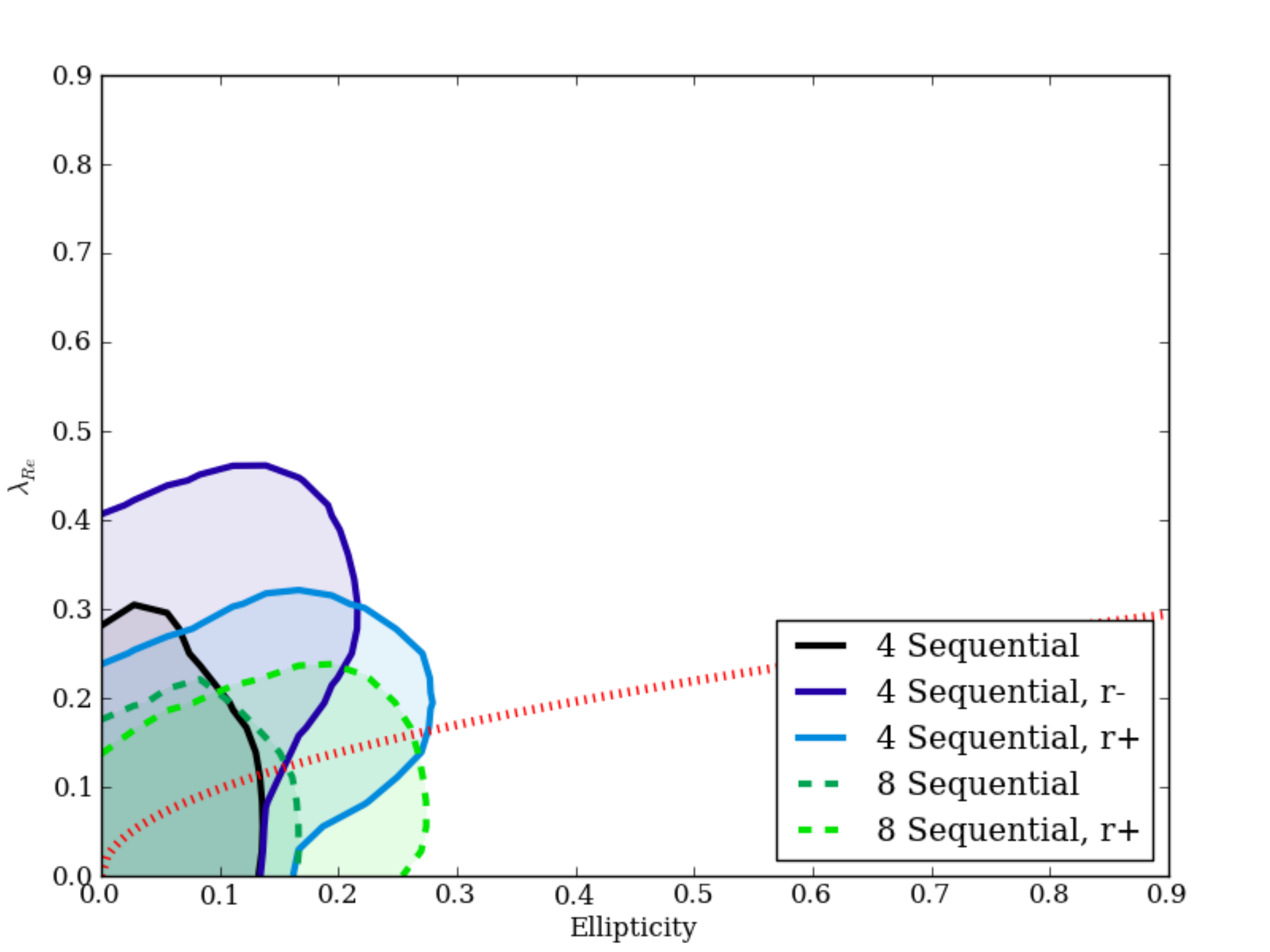,height=7cm}
	\caption{Same as Figure \ref{le-remerger}, but showing multiple sequential merger simulations with either four or eight identical G1 or G2 progenitors. Sequential simulations begin with four or eight progenitors, and have a staggered formation history. In contrast to the binary merger tree simulations, the merger ratio decreases from the initial 1:1 value to 1:3 (1:7) in the four (eight) progenitor case. Increasing the number of progenitors in sequential simulations from 4 to 8 yields still rounder and slower remnants. Remnants formed via sequential merging form slow and round rotators representative of the slow rotators found in the \atlas3d\ observations.}
	\label{le-sequential}
\end{figure}

Figure \ref{le-remerger} shows the remnants of binary merger tree simulations, ranging from four progenitors merged in two rounds, to eight progenitors merged in three rounds of major merging. The remnants of all four-progenitor binary merger tree simulations are fast rotators, with most projections having $\lambda_R>0.31\sqrt{\epsilon}$ and significant flattening. Increasing the number of progenitors to eight and thereby adding another round of merging yields remnants with largely the same \lamR\ but lower $\epsilon$ than remnants formed from two rounds of merging. 

Using the Millenium Simulation and semi-analytic models, \citet{2006MNRAS.366..499D} found that systems with stellar mass $M_* < 10^{11} M_{\odot}$ formed from fewer than two massive progenitors, but rising to five progenitors for the most massive systems. In the \atlas3d\ sample, these most massive systems are predominantly slow rotators, suggesting that multiple mergers may give rise to slow rotators. Simulations of multiple minor mergers demonstrate that a transformation from a spiral progenitor into an elliptical galaxy takes place when the cumulative mass added exceeds 30\%-40\% \citep{2007A&A...476.1179B}. With increasing numbers of mergers, the overall $V/\sigma $ dropped and remnant systems became progressively more round. Nevertheless, remnants comparable to the \atlas3d\ slow rotators were not found, and in this paper we test whether further merging would create a suitably slow remnant. 

Figure \ref{le-sequential} shows the remnants of the sequential merging of four progenitors. Four-progenitor simulations yield remnants that are more slowly rotating than any binary merger in our suite. For this series, the mean \lamR\ varies from 0.03 to 0.12, with the mean ellipticity ranging from 0.1 to 0.2. Remnants of sequential merging are slower and rounder than those produced in binary tree merging. As in the binary merger tree case, doubling the number of progenitors does not decrease the ellipticity but does decrease the rotational support (i.e. the mean \lamR).

In summary, compared with multiple mergers, binary mergers have fast rotator remnants with higher ellipticities and in general fall on the top-right of the diagram. Individual minor merger events keep much of the progenitor structure intact and preserve the high rotational velocities, and as a result are the fastest rotators in our sample. Remnants of major mergers have slightly less rotational support when compared to minor merger remnants, and form the second-fastest rotators of our sample. Varying the gas fraction in progenitors does not produce any slow rotators, but does yield the largest range of both \lamR\ and $\epsilon$ values of any other category of mergers studied here. The poorest gas fraction simulation produces remnants with low \lamR\ and low $\epsilon$, which is similar to simulations of dissipationless binary mergers \citep{2006ApJ...650..791C}. The most gas-poor mergers result in slowly-rotating but highly elongated systems. Variations in the pericenter distance or orbital ellipticities do not generate slow rotators, but does produce a spread of photometric ellipticities. Finally, multiple mergers lead to the slowest, roundest rotators. 
Binary merger tree simulations with either four or eight progenitors produce remnants that, while quite round, are much faster than sequential mergers. In the majority of projections, these remnants are above $\lambda_R=0.31\sqrt{\epsilon}$ and thus classified as fast rotators.

\begin{figure*}
	\vbox to 110mm{\vfil
	\psfig{file=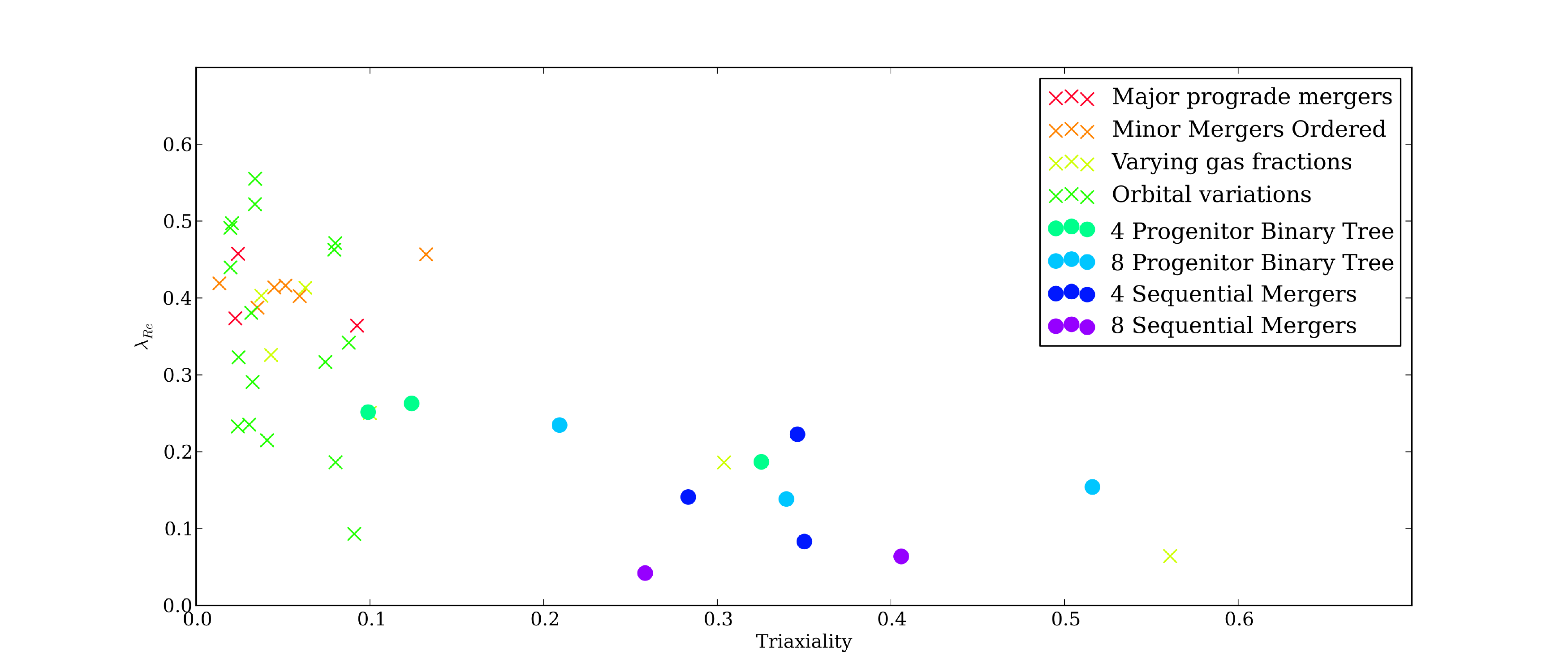,height=80mm}
    \caption{ $\lambda_R$ versus the triaxiality evaluated at $1.0$~\Reff, and averaged over each projection of a simulation. As in Figure \ref{lamellip}, filled circles represent multiple merger simulations, and cross marks are binary major mergers. Multiple mergers tend to produce triaxial slow rotators while binary mergers lead to largely fast rotators with little triaxiality.}
	\vfil
\label{lam_triax}
}
	
\end{figure*}

Sequential mergers with eight progenitors produce slower and rounder remnants than four progenitors, 
which agrees with the current conventional wisdom that multiple minor mergers are the most likely mechanism for producing this class of galaxies (e.g., \citealt{2011MNRAS.417..845K}).
We note that our merging sequences do not extend fully into the minor-dominated merger regime that may be typical for
massive galaxies in dense environments: our eight-progenitor merger mass-ratio (both number-weighted and mass-weighted) is 1:2.7, 
while cosmological simulations suggest $\sim$\,1:10 (mass-weighted) or $\sim$\,1:40 (number-weighted) may be typical for this class of galaxies \citep{2013arXiv1311.0284N}.
Given the general trends from our controlled experiments, we expect that such mergers, if fairly isotropic, could produce even better agreement with the 
observed class of round, slow rotators -- with our experiments perhaps representative of the first stages of their formation.


\section{Triaxiality and twists}\label{sec:kin}

In a series of simulations, \citet{Jesseit:2009p103} showed that fast rotators tend to have lower triaxialities, and slow rotators have higher triaxialities. We revisit this question with knowledge of the initial merger conditions (see Figure \ref{lam_triax}). Binary mergers yield fast rotating but non-triaxial remnants. As demonstrated by \citet{2006ApJ...650..791C}, the triaxiality depends strongly on the initial orbital parameters of the progenitor disk galaxies, which leads to the scatter in triaxiality from $0.0-0.1$ in our binary merger remnants. The two lowest gas fraction initial conditions are the only two binary merger remnants to attain high triaxialities. Sequential merger simulations on the other hand are generally triaxial and much more slowly-rotating. Binary merger tree simulations yield mildly triaxial galaxies with $T$ ranging from $0.1-0.3$ for four progenitors and climbing to $0.2-0.5$ for eight progenitors. Remnants of four sequential mergers have higher triaxiality than the binary tree mergers, and doubling the number does not appear to influence the shape of the remnant dramatically, while still decreasing the overall rotation.

In this section we report the incidence of kinematic twists and kinematically decoupled cores for various classes of mergers. While the majority of the projected LOS quantities display regular rotation chracterized by a rotational velocity that climbs smoothly with radius, many kinematic maps have distorted features. To study how these features arise, we correlate the incidence of kinematic twists and decoupled cores with the formation histories of the merger remnants.
\begin{figure}
	\psfig{file=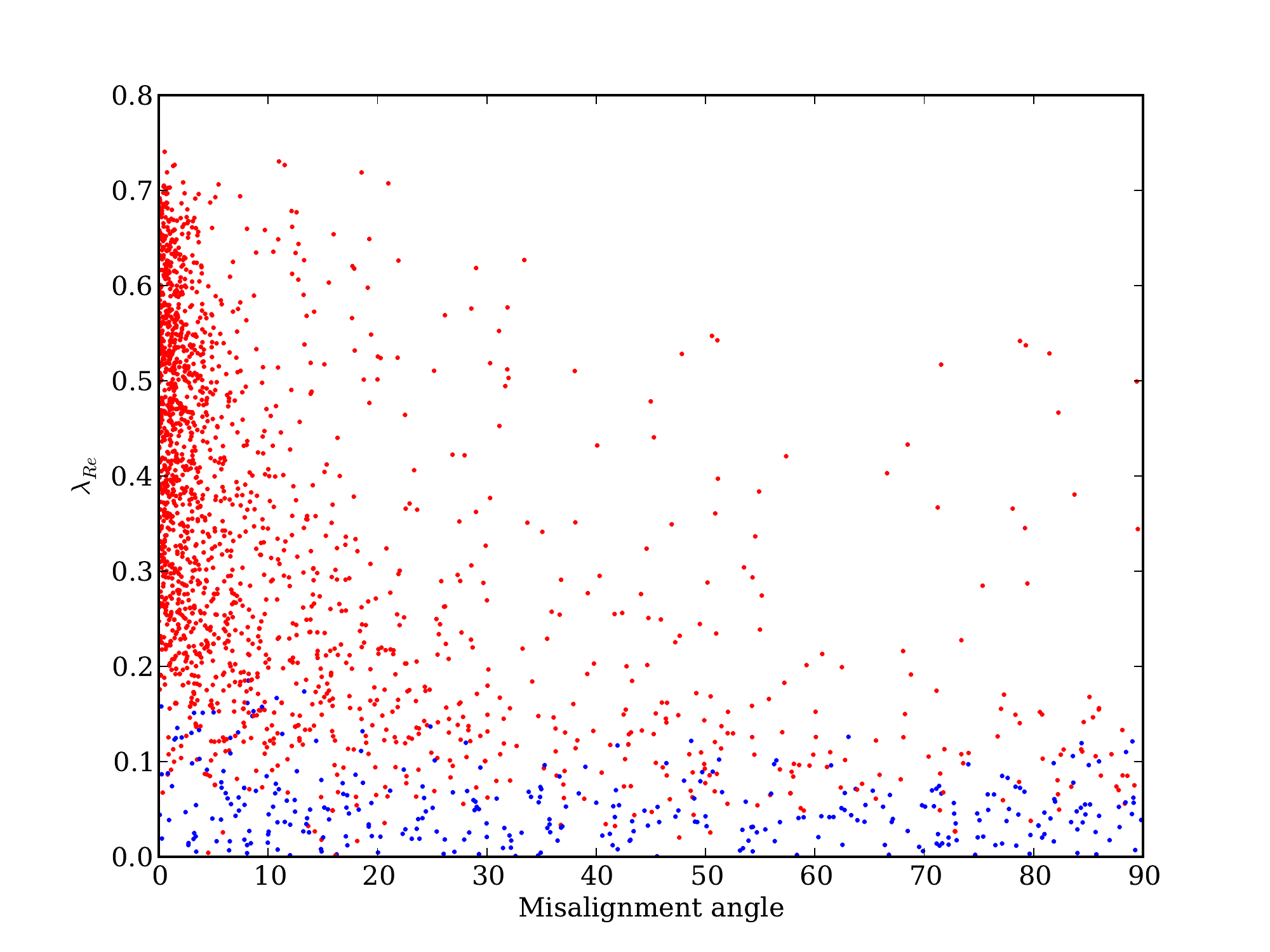,height=7cm}
	\caption{$\lambda_{Re}$ and the kinematic misalignment evaluated at $1.0$~\Reff\ shown for each projection for all simulations. Fast rotators, shown in red, satisfy $\lambda_{R_e} \geq 0.31\sqrt{\epsilon}$; slow rotators are shown in blue. For readability, only 10$\%$ of points are shown. Fast rotators tend to have highly coupled kinematic and photometric axes with misalignments typically less than 10$^{\circ}$. Below $\lambda_R=0.1$, most projections are slow rotators with no alignment between the kinematic and photometric axes, and hence have misalignments ranging from 0 to $90^{\circ}$.
	\label{l_psi}
}
\end{figure}

\begin{figure}
	\psfig{file=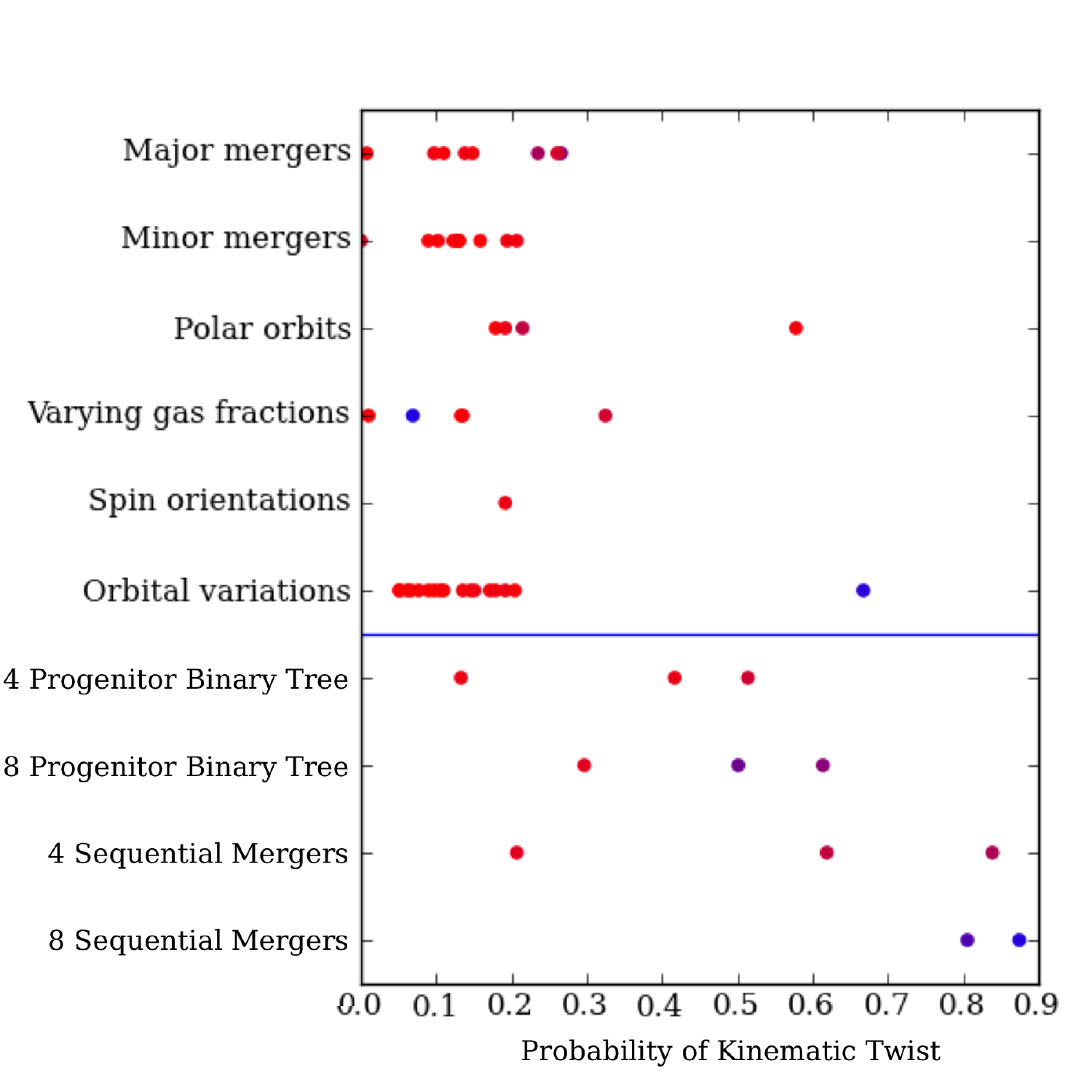,width=\columnwidth}
    \caption{Kinematic twists are defined as a gradual change in the kinematic angle that changes by at least $10^{\circ}$ over all radii. Points represent the probability of observing a kinematic twist in various categories of merger scenarios. Blue dots are shown for simulations which are predominantly slow rotators,while red are fast rotators. Roughly speaking, binary mergers typically have a $15\%$ chance of finding a KT. The two slow rotators formed in the binary mergers have dramatically different behaviors. The gas-poor slow rotator forms almost no KTs but the slow rotator formed with an orbital variation of zero initial angular momentum has a high ($70\%$) probability of finding a KT. Multiple merger simulations generally host KTs with much greater frequency than binary merger simulations, with sequential mergers with eight progenitors having $80\%-90\%$ chance of finding a KT in any single projection.}
	\label{kc_kt}
\end{figure}

Figure \ref{kc_kt} demonstrates that binary mergers typically have KTs in $10\%-20\%$ of projections, while multiple mergers produce higher probabilities of KTs. There are two exceptional binary mergers where the probability of finding a KT is $\approx 60\%$ .  The first is a polar orbit where the angular momenta of the two progenitors are nearly equal in magnitude but perpendicular in direction, which yields a fast rotator with high incidence of KTs. Out of the full set of polar orbit set of simulations, this simulation maximizes the initial misalignment in orbital angular momenta, and has the maximum incidence of KTs in the remnant. The second binary merger with a high incidence of KTs occurs in a unique orbital variation where the orbits are initialized with a net angular momentum of zero. By construction, this forms a slowly-rotating remnant with high incidence of KTs.

The incidence of KTs in multiple mergers is in general much higher and spans a larger range than in binary mergers. Binary merger tree simulations have KT incidences spanning $15\%-60\%$, while sequential mergers are much more likely to be slower rotators with more KTs. In the most extreme case, eight sequential mergers yields a twist in nearly $90\%$ of all projections. In both binary merger tree and sequential simulations, doubling the number of progenitors increases the likelihood of finding a KT. Unlike binary mergers, the likelihood of finding a slow rotator appears to scale with the chance of finding a KT.

\begin{figure}
		\psfig{file=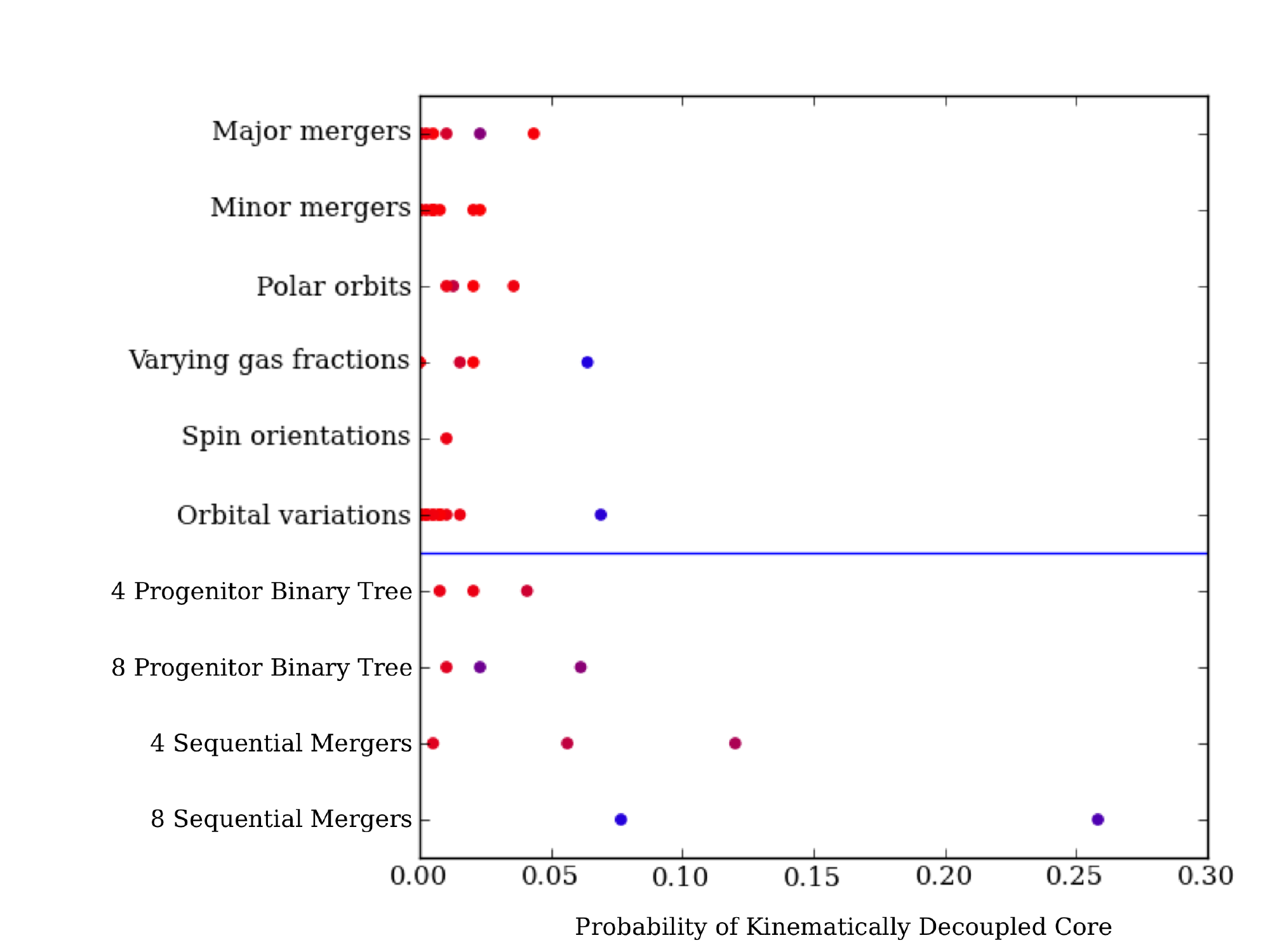,width=\columnwidth}
	\caption{Same as Figure \ref{kc_kt} but with the incidence of kinematically decoupled cores. To be labeled a KDC, the kinematic moment must undergo at least a change of $30^\circ$ in the orientation of the kinematic moment in a single radial bin. Above the blue line are binary mergers and below are multiple merger simulations. KDCs are more commonly found in sequential merger simulations than in binary mergers. Within any single category of simulations, slowly rotating remnants are the more likely to host a KDC than fast rotators.}
	\label{kc_kdc}
\end{figure}

Figure \ref{kc_kdc} shows that binary merger remnants rarely host kinematically decoupled cores, with KDCs occurring only in $1\%-5\%$ of projections. Slowly rotating binary remnants feature KDCs, with somewhat increased frequency when compared to fast rotators. Binary merger tree simulations have similarly low incidences of KDCs, although the most slowly-rotating remnants do not appear to have an increased incidence of decoupled cores. Sequential mergers have the highest rate of decoupled cores.  The slow-rotator remnants of eight sequential mergers are more likely to feature a KDC than any binary merger remnant in our sample.

\begin{figure}
	\psfig{file=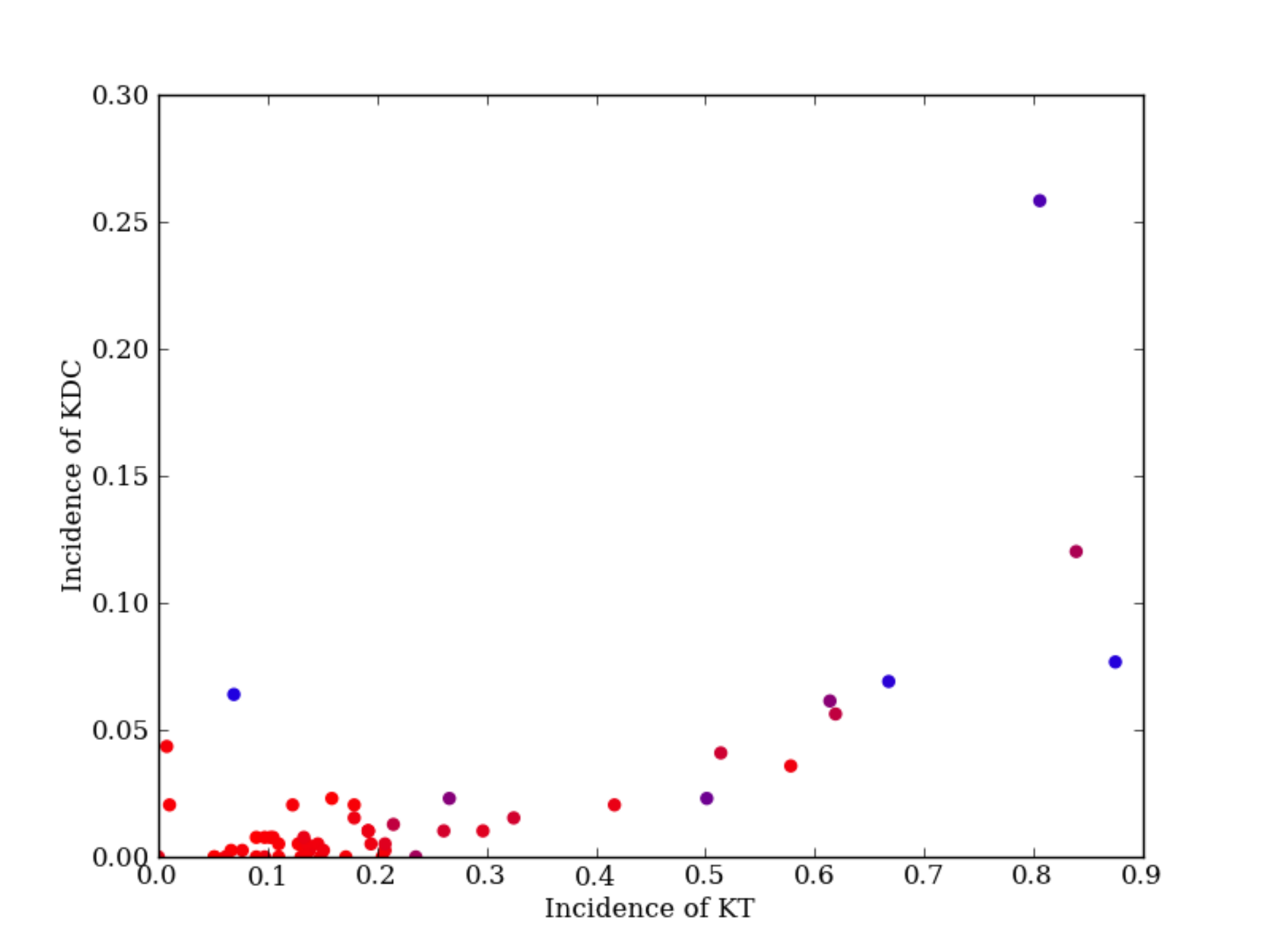,height=7cm}
	\caption{For all simulations, the probability of finding a KT is compared to the probability of finding a KDC in the same simulation. Slow rotators are colored blue, fast rotators red. For simulations with KT rates above $50\%$, which are predominantly multiple merger simulations, KTs correlate with KDCs, suggesting that these two phenomena are linked and that KDCs are extreme forms of KTs. }
	\label{kc_kt-kdc}
\end{figure}

Simulations featuring KTs in $0-30\%$ of projections have a constant rate of KDCs. Once the majority of projections feature a twist, the rate of decoupled cores also scales with the KT rate. However, there are only two binary merger simulations with twisting rates higher than $50\%$, so the trend could be the result of differences between multiple and binary mergers instead of an intrinsic correlation between KTs and KDCs.


\section{Summary}\label{sec:sum}
In this paper we present a kinematic analysis and kinematic classification of 95 simulated remnants spanning a range of initial conditions. Our simulated binary merger remnants vary in mass ratio, orbital pericenter, orbital ellipticity, spin, and gas fraction. We also include multiple merger simulations where the remnant is grown exclusively through major merging or where the progenitor grows through sequential and increasingly more minor merging.

Nearly all of our binary merger remnants resemble \atlas3d\ fast rotators. These simulations have a small chance of being observed as a slow rotator ($\approx5\%$ of projections). \lamR\ rises steadily with $\epsilon$, the mean photometric-kinematic moments are aligned to within $5\%$, and the stellar components are well-described by an oblate ellipsoid. We find that a single major or minor merger does not yield a slowly-rotating and spherical remnant. However, low gas fractions of $\approx 10\%$ lead to merger remnants that are slowly-rotating but elongated, a result qualitatively similar to simulations of dissipationless systems \citep{2006ApJ...650..791C}. Simulations where the progenitors' initial angular momenta and orbital angular momentum sums to zero also lead to remnants that are slow rotators but are nevertheless still highly elliptical. Thus, even in binary mergers simulations that lead to the formation of slowly rotating remnants, the remnants are highly elongated and largely incompatible with the slow rotators as found by \atlas3d.

In contrast to binary merger simulations, remnants of multiple mergers resemble observed slow rotators. We study two classes of multiple merger histories, both having progenitors identical in number and properties, but differ in the order that they are merged. `Binary merger tree simulations' form a remnant exclusively through generations of binary major merging, and `sequential' multiple mergers form a remnant by multiple, largely non-overlapping, increasingly minor mergers. Broadly speaking, both binary tree merger remnants and sequential merging remnants were rounder and slower than binary merger remnants. Thus we find that the merger history is an important ingredient to forming slow or fast rotators (see also \citealt{Navarro:2013}). 
Despite having identical progenitors as binary merger tree simulations, the sequential merger remnants were more likely to be round slow rotators. For both the binary merger tree and sequential multiple merger simulations, increasing the number of progenitors from four to eight yielded remnants that were rounder and more slowly-rotating. Our results suggest that multiple mergers that grow exclusively through generations of major merging yield predominantly fast-rotators, but growth through minor merging largely yields slow mergers.  

Kinematic twists are found ($\sim$10-20\% of projections) in our binary mergers, but are much more common ($\sim$30-80\%) in remnants of multiple mergers. Kinematically decoupled cores are found infrequently in binary mergers ($\approx5\%$) but multiple merger remnants commonly host KDCs (5\%-30\%). Kinematically distinct cores are overall less prevalent than KTs, but again are most commonly found in multiple merger remnants, particularly in sequential multiple merging. Overall, we find that the incidence of kinematic twists holds regardless of whether in a given projection the remnant is a slow or fast rotator.

In conclusion, our controlled experiments support the emerging framework that massive, round, slow-rotating ellipticals can be formed only through the accumulation of many minor mergers, while fast rotators may be formed through a variety of pathways.  More work is now needed to connect the infall patterns of galaxies in a full cosmological context with their rotational outcomes.  Further clues to the specific merging conditions for both fast and slow rotators could come through more detailed analyses of their kinematics, including in the outer regions (e.g. \citealt{Schauer:2014,Arnold:2014,Wu:2014,Raskutti:2014}).


\section{Acknowledgments}

We thank Frederic Bournaud for a helpful report, and Michele Cappellari, Duncan Forbes and Dan Taranu for comments.
This work was supported by National Science Foundation grant AST-0909237 and by computing resources at the NASA Ames Research Center.
G.S.N. was supported by the Department of Energy Computational Science
Graduate Fellowship administered by the Krell Institute and by the ERC
European Research Council under the Advanced Grant Program Num
267399-Momentum. 

\bibliographystyle{mn2e}
\bibliography{bib}

\end{document}